\documentclass[%
 reprint,
superscriptaddress,
 amsmath,amssymb,
floatfix,
]{revtex4}
\usepackage{graphicx}
\usepackage{dcolumn}
\usepackage{bm}
\usepackage{algorithm}
\usepackage{algpseudocode}
\usepackage{amsmath}
\usepackage{braket}
\usepackage{multirow}
\usepackage{hyperref}
\hypersetup{
colorlinks=true, citecolor = blue, linkcolor=blue, filecolor=blue
}
\usepackage{url}
\usepackage{color,xcolor}
\usepackage{comment}
\usepackage{booktabs} 
\newcommand{\tabincell}[2]{\begin{tabular}{@{}#1@{}}#2\end{tabular}}
\begin{document}
\preprint{APS/123-QED}

\title{Observation of freezing phenomenon in high-dimensional quantum correlation dynamics}

\author{Yue Fu}
\thanks{These authors contributed equally to this work.}
\affiliation{CAS Key Laboratory of Microscale Magnetic Resonance and School of Physical Sciences, University of Science and Technology of China, Hefei 230026, China}
\affiliation{School of Integrated Circuits and Electronics, Beijing Institute of Technology, Beijing 100081, China}

\author{Wenquan Liu}
\thanks{These authors contributed equally to this work.}
\affiliation{Institution of Quantum Sensing, Zhejiang University, Hangzhou 310027, China}

\author{Yunhan Wang}
\affiliation{CAS Key Laboratory of Microscale Magnetic Resonance and School of Physical Sciences, University of Science and Technology of China, Hefei 230026, China}
\affiliation{Hefei National Laboratory, University of Science and Technology of China, Hefei 230088, China}

\author{Chang-Kui Duan}
\affiliation{CAS Key Laboratory of Microscale Magnetic Resonance and School of Physical Sciences, University of Science and Technology of China, Hefei 230026, China}
\affiliation{Hefei National Laboratory, University of Science and Technology of China, Hefei 230088, China}
\affiliation{CAS Center for Excellence in Quantum Information and Quantum Physics, University of Science and Technology of China, Hefei 230026, China}

\author{Bo Zhang}
\affiliation{
School of Physics, Beijing Institute of Technology, Beijing 100081, China
}

\author{Yeliang Wang}
\affiliation{School of Integrated Circuits and Electronics, Beijing Institute of Technology, Beijing 100081, China}

\author{Xing Rong}
\email{xrong@ustc.edu.cn}
\affiliation{CAS Key Laboratory of Microscale Magnetic Resonance and School of Physical Sciences, University of Science and Technology of China, Hefei 230026, China}
\affiliation{Hefei National Laboratory, University of Science and Technology of China, Hefei 230088, China}
\affiliation{CAS Center for Excellence in Quantum Information and Quantum Physics, University of Science and Technology of China, Hefei 230026, China}

\begin{abstract}
Quantum information processing (QIP) based on high-dimensional quantum systems provides unique advantages and new potentials where high-dimensional quantum correlations (QCs) play vital roles. 
Exploring the resistance of QCs against noises is crucial as QCs are fragile due to complex and unavoidable system-environment interactions. 
In this study, we investigate the performance of high-dimensional QCs under local dephasing noise using a single nitrogen-vacancy center in diamond. 
A freezing phenomenon in the high-dimensional quantum discord dynamics was observed, showing discord is robust against local dephasing noise.
Utilizing a robustness metric known as freezing index, we found that the discord of qutrits outperforms their qubits counterpart when confronted with dephasing noise.
Furthermore, we developed a geometric picture to explain this intriguing freezing phenomenon phenomenon.
Our findings highlight the potential of utilizing discord as a physical resource for advancing QIP in high-dimensional quantum settings.
\end{abstract}

\maketitle

\section*{Introduction}
Quantum correlations (QCs), a major feature of quantum mechanics that cannot be explained by classical theory, play essential roles in quantum information processing (QIP) with enhanced performances \cite{review_2009,review_2020_Uola, JPhys_2016_Adesso,book_2015_streltsov,review_2018_Braun}.
However, due to the interactions with the environment, quantum systems are inevitably affected by various noises \cite{review_2016_Suter}, which cause degradation of QCs \cite{review_2019_Lewis-Swan} and failure of QIP tasks.
Several methods have been proposed and demonstrated to mitigate this problem, including dynamical decoupling \cite{NC_2021_wang}, decoherence-free subspace \cite{PRL_2023_zhang, PRL_2024_wang}, error correction codes \cite{NC_2023_Sundaresan}, etc.
Meanwhile, exploring noise-robust quantum featurescite \cite{PRL_2010_Mazzola, PRA_2011_Karpat, PRL_2015_Carnio} and designing QIP algorithms based on them may be a distinct pathway to deal with the noise issue \cite{JPhys_2016_Adesso}.
For example, it has been found that quantum discord, a kind of characterization for QCs \cite{ PRL_2001_Olivier, JPA_2001_Henderson}, can be frozen for a period when suffering dephasing noise \cite{PRL_2010_Mazzola, review_2012_Modi, review_2018_De, review_2018_Bera, NP_2012_Dakić}.
This intriguing phenomenon shows the immunity of discord against the dephasing noise and has been studied intensively in two-level qubit systems \cite{NC_2013_Xu, PRL_2011_Auccaise, NC_2010_Xu, EPL_2017_Singh}.

Recently, with the rapid advances in quantum technologies, there have been increasing explorations of utilizing high-dimensional quantum systems to execute QIP tasks \cite{AQT_2019_Cozzolino, FP_2020_Wang,review_2020_Erhard,review_2023_Guo}.
And thanks to the exponential expansion of the Hilbert space, high-dimensional quantum systems presents unique advantages \cite{PRX_2023_Liu, Science_2018_Wang, NC_2023_Hrmo} and new potentials \cite{NC_2024_Fernández,  NC_2022_Chi, NP_2022_Ringbauer, NP_2019_Reimer} over their qubit counterparts in QIP.
However, the noise issue remains and can be more sticky \cite{PRA_2015_Vitanov, NC_2020_Coladangelo}.
Take the dynamical decoupling method as an example, the increased dimension makes the design and implementation of decoupling sequences more complicated \cite{PRR_2021_Napolitano, PS_2022_Singh, PRL_2018_Kraft}.
And up to six $\pi$ pulses were performed to realize dynamical decoupling of single qutrit in a recent experiment \cite{PRA_2022_yuan}.
In this regard, exploring noise-robust quantum features may be a promising method \cite{IJQI_2022_xiao, SR_2017_Cárdenas, PRL_2022_Fu,PRA_2010_Ali}. 
However, investigations into the dynamics of high-dimensional quantum correlations have so far been limited \cite{PRA_2009_Maziero,PRA_2009_Werlang,PRL_2010_Mazzola, PRL_2010_Lang, PRA_2015_Chanda}.

In this work, we investigate the dynamics of various quantum correlations between two qutrits under the one-qutrit local dephasing noise, which is common in quantum systems composed by different spins. 
We first proposed a geometric picture to show that the dynamics of high-dimensional quantum discord could be frozen and exhibit an intriguing sudden transition phenomenon for a type of Bell-diagonal states.
Then, working on two qutrits consisted by the electron spin and the nuclear spin of a single nitrogen-vacancy (NV) center in diamond \cite{PhysRep_2013_Doherty}, this dynamical property was confirmed experimentally. 
Finally, we compared the goodness of this freezing behavior between qutrits and qubits utilizing a measure called freezing index \cite{PRA_2015_Chanda} since there exhibits a certain trade-off between the amount of discord and the time that it could be frozen \cite{Review_2017_Cham}.
The results show that qutrits possess a larger freezing index than qubits, demonstrating its potential in future high-dimensional QIP.

 \begin{figure}\centering
	\includegraphics[width=0.95\columnwidth]{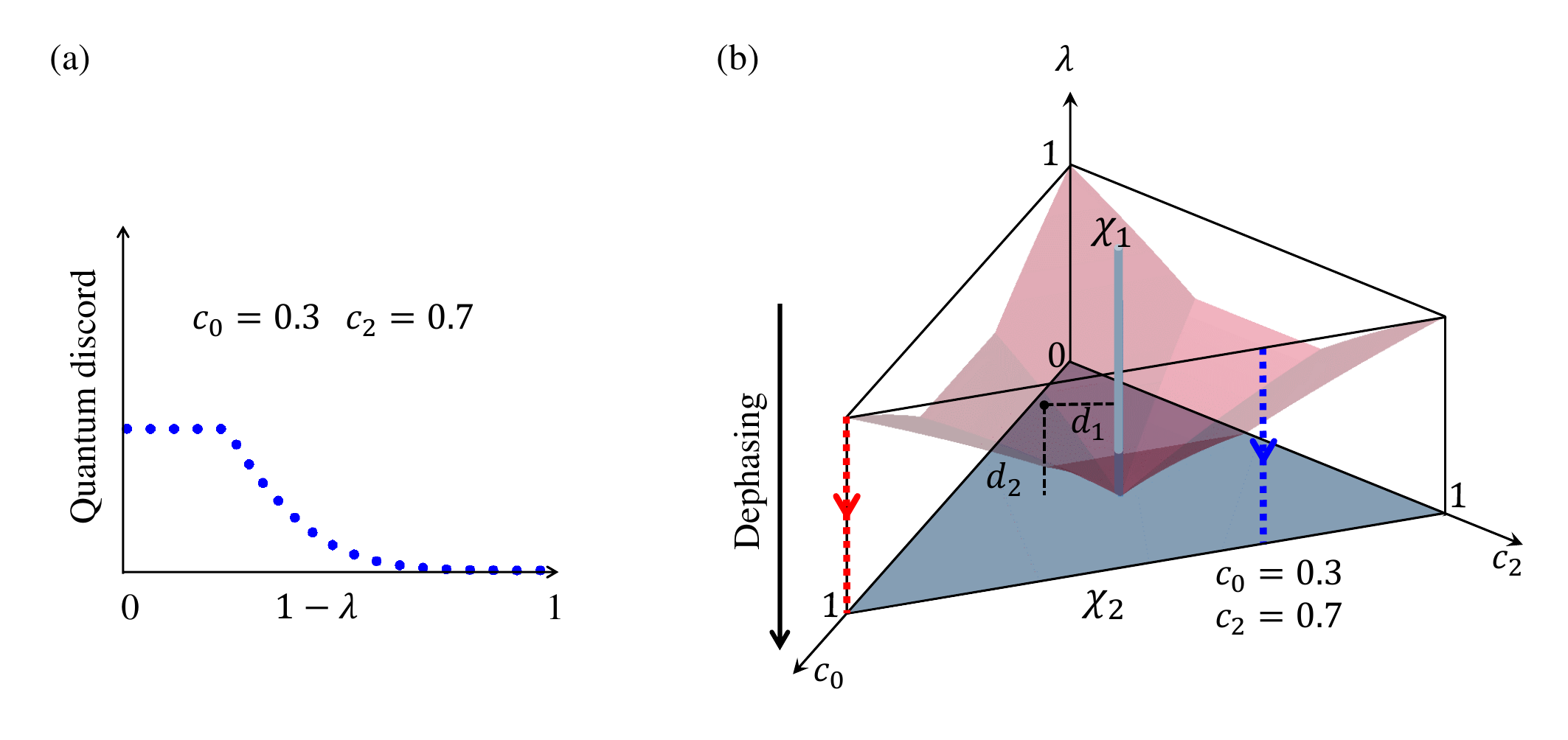}
	\caption{\textbf{QC dynamics of the Bell-diagonal states $\rho_{\rm BD,3}$ under one-qutrit local dephasing noise.}
		(a) The quantum discord presents a freezing phenomenon.
		(b) Geometric perspective of the freezing phenomenon.
		$c_0,c_2\in [0,1]$ and $c_0+c_2\le1$ are the parameters of $\rho_{\rm BD,3}$, $\lambda\in [0,1]$ characterizes the dephasing process.
		These parameters construct a parameter space in the form of a triangular prism.
		The gray line $\chi_1$ and plane $\chi_2$ are the states with zero quantum discord.
		For other quantum states, their minimal distance to $\chi_1$ and $\chi_2$ are defined as $d_1$ and $d_2$, respectively, and the quantum discord is min$\{d_1, d_2\}$.
		The pink surface shows the states with $d_1=d_2$.
		If the dephasing trajectory crosses the pink surface (blue dash line for example), the quantum discord dynamics will present a sudden transition.}	
	\label{Fig1}
\end{figure}

\section*{Theory}
For a bipartite quantum system, the quantum discord between the two subsystems is defined as the minimum Schatten 1-norm distance between the system state $\rho_{AB}$ and state $\sigma_{AB}$ within the set of quantum-classical states $\{\rho^{\rm q-c}\}$ that has zero discord:
$\mathcal{Q}=  \underset{\sigma_{AB}\in\{\rho^{\rm q-c}\}}{{\rm min}} {\rm } ||\rho_{AB}-\sigma_{AB}||_1$ \cite{PRL_2010_Dakić,EPL_2013_Paula, PLA_2016_Jakóbczyk}.
Here, we consider a family of two-qutrit Bell-diagonal states. They are the superposition of maximally entangled states given by \cite{SR_2017_Cárdenas}
\begin{align}
    \rho_{\rm BD,3}\! =\! c_0 |\Psi_{00}\rangle \langle \Psi_{00}|\!+\!c_1 |\Psi_{01}\rangle \langle\Psi_{01}|\!+\!c_2|\Psi_{02}\rangle \langle \Psi_{02}|,
    \label{BD_States}
\end{align}
with $|\Psi_{00}\rangle = (|{00}\rangle+|11\rangle+|{22}\rangle)/\sqrt{3}$,  $|\Psi_{01}\rangle =(|{01}\rangle+|12\rangle+|{20}\rangle)/\sqrt{3}$ and $|\Psi_{02}\rangle = (|{10}\rangle+|21\rangle+|{02}\rangle)/\sqrt{3}$ in the computational basis. Parameters $c_0,c_1,c_2\in [0,1] $ and satisfy $c_0+c_1+c_2 = 1$.
The evolution of these states under the local dephasing noise of the first qutrit can be written as
\begin{footnotesize}
    \begin{align}
    \rho_{\rm BD,3}^{\rm deph}\!(c_0,\!c_2,\!\lambda)\!\!=\!\!\frac{1}{3}\!\!\left(\!\!\!\begin{array}{ccccccccc}
    			c_0\!&\!  0\!&\!  0\!&\!  0\!&\!c_0\lambda \!&\!  0\!&\!  0\!&\!  0\!&\!c_0  \lambda\!^2\\
    			0\!&\!  c_1\!&\!  0\!&\!  0\!&\!    0\!&\!c_1\lambda\!&\!c_1\lambda\!^2\!&\!    0\!&\!   0\\
    			0\!&\!  0\!&\!c_2  \!&\!c_2\lambda\!&\!   0\!&\!    0\!&\!    0\!&\!c_2\lambda\!^2\!&\!   0\\
    			0\!&\!    0\!&\!c_2\lambda\!&\!c_2\!&\!    0\!&\!    0\!&\!    0\!&\!c_2\lambda\!&\!   0\\
    			c_0\lambda\!&\!    0\!&\!    0\!&\!    0\!&\!c_0\!&\!    0\!&\!    0\!&\!    0\!&\!c_0\lambda  \\
    			0\!&\!c_1\lambda\!&\!    0\!&\!    0\!&\!    0\!&\!c_1\!&\!c_1\lambda\!&\!    0\!&\!   0\\
    			0\!&\!c_1\lambda\!^2\!&\!    0\!&\!    0\!&\!    0\!&\!c_1\lambda\!&\!c_1\!&\!    0\!&\!   0\\
    			0\!&\!   0\!&\!c_2\lambda\!^2\!&\!c_2\lambda\!&\!    0\!&\!    0\!&\!    0\!&\!c_2\!&\!   0\\
    			c_0\lambda\!^2 \!&\!    0\!&\!    0\!&\!    0\!&\!c_0\lambda\!&\!    0\!&\!    0\!&\!    0\!&\!c_0
    \end{array}\!\!\!\right)\!\!,
    \label{Dephased_BD_States}
    \end{align}
\end{footnotesize}
where $\lambda = \exp{\left[-(t/T_2^*)^n\right]}$ denotes the decay of the off-diagonal elements with $T_2^*$ characterizing the dephasing of the first qutrit and $n$ being the noise-dependent stretch factor.
Figure \ref{Fig1}(a) shows a typical discord dynamics for quantum states with the form shown in Eq. \ref{Dephased_BD_States}. 
Under the local dephasing noise, the discord can be frozen for a period before a sudden transition happens and the discord decays to zero gradually \cite{SR_2017_Cárdenas}.
This phenomenon shows that discord is robust against the dephasing noise.
The freezing phenomenon was first discovered in qubit systems where corresponding theoretical explanations and experimental observations have been explored intensively.
Below we analyze the model in detail and geometrically elucidate the origin of this exotic phenomenon in the qutrit case.

As shown in Fig. \ref{Fig1}(b), the state $\rho_{\rm BD,3}^{\rm deph}$ is specified by 3-parameters $c_0,c_2$ and $\lambda$.
The matrix is physical when $c_0+c_2\le1$, corresponds to the region of the triangular prism.
We derived and calculated the set of quantum-classical states with zero discord and found it contains two parts:
the gray line $\chi_1$ where $c_0=c_2=1/3$, and the gray plane $\chi_2$ where $\lambda = 0$ (see supplemental material for details).
For a point in the parameter space that corresponds to a quantum state, $d_1$ and $d_2$ are utilized to denote the minimal distance of it to $\chi_1$ and $\chi_2$, respectively.
The discord of this state is ${\rm min}\{d_1, d_2\}$.
The states that satisfy $d_1=d_2$ are plotted by the pink surface in Fig. \ref{Fig1}(b).
Equation of this surface is given in the supplementary information.

	\begin{figure}\centering
		\includegraphics[width=1\columnwidth]{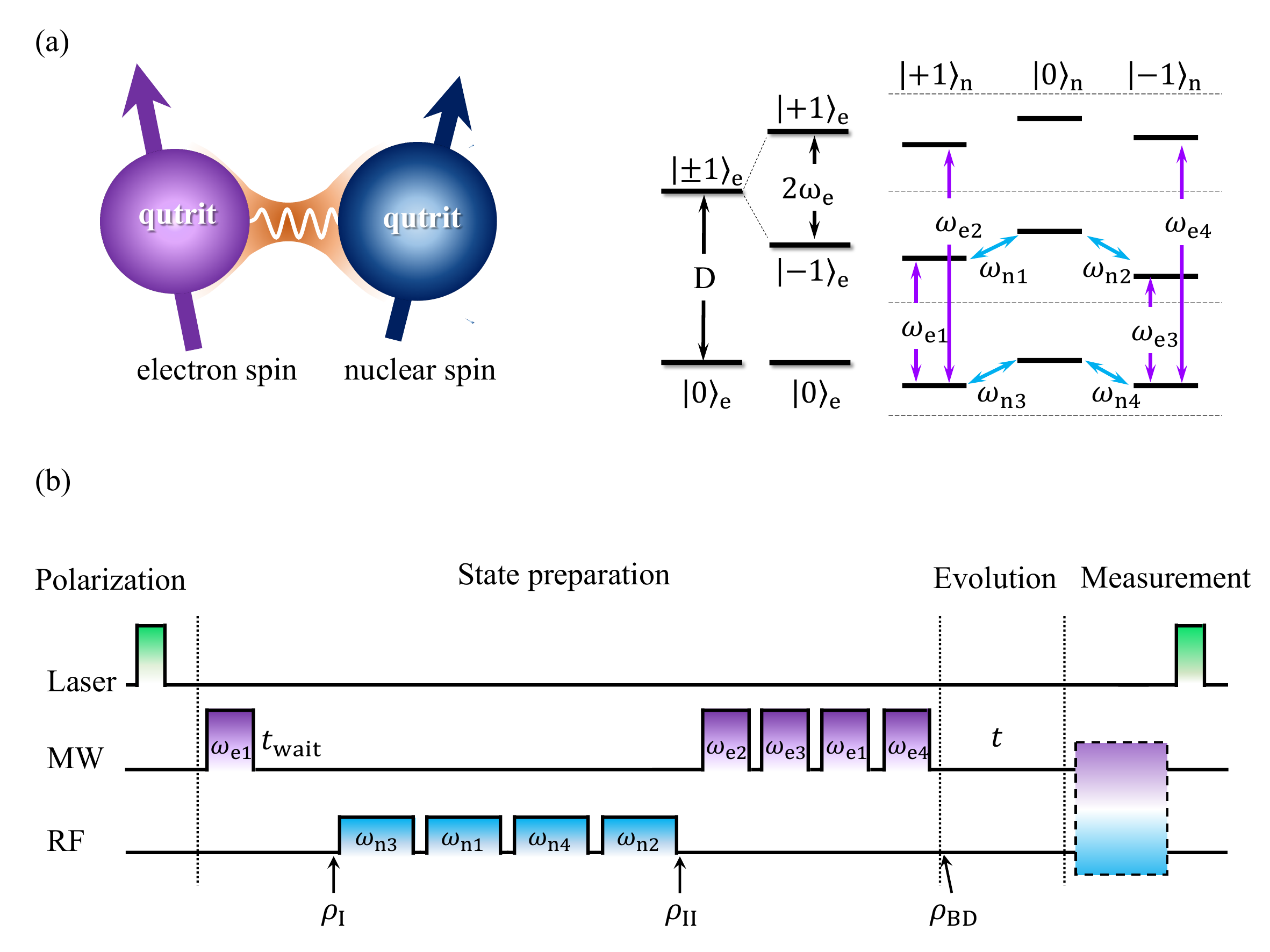}
		\caption{\textbf{The NV center two-qutrit system and the experimental pulse sequence diagram.}
			(a) Left: in the ground state of the NV center, the electron spin and the nuclear spin constitute a two-qutrit system.
			Right: the energy levels. Transitions between different electron (nuclear) spin states can be controlled by microwave (radio frequency) pulses indicated by purple (blue) arrows.
			(b) Polarization, state preparation, evolution under dephasing noise and measurement are displayed.
			In the state preparation part, selective MW (purple) and RF (blue) pulses were performed to generate the Bell-diagonal state $\rho_{\rm BD,3}$.
			The measurement part comprises various sequences to read out different elements of the density matrix (see  supplemental material for details).
		}
		\label{Fig2}
	\end{figure}

The sudden transition of the discord dynamics can be explained as follows.
The evolution trajectory of $\rho_{\rm BD,3}$ with given $c_0$ and $c_2$ under the one-qutrit local dephasing noise is a straight line that parallels the $\lambda$-axis.
The blue dashed line in Fig. \ref{Fig1}(b) with
$c_0=0.3$ and $c_2=0.7$ is an example, which corresponds to the one used in Fig. \ref{Fig1}(a).
The beginning part of this trajectory is above the pink surface, where $d_1<d_2$ and $\mathcal{Q}=d_1=\sum_{i=0}^{2}|c_i-1/3|$ (see  supplemental material).
The independence of $d_1$ on $\lambda$ makes discord a constant in this part, corresponding to the freezing interval shown by the discord dynamics in Fig. \ref{Fig1}(a).
After the trajectory crosses the pink surface, we have $d_1>d_2$ and $\mathcal{Q}=d_2=\lambda^2+\sqrt{\lambda^4+8\lambda^2}$.
Thus the discord decays to zero smoothly as $\lambda$ approaches zero gradually.
The discord dynamics of other trajectories can be analyzed similarly.
In general, the sudden transition will happen whenever a dephasing trajectory crosses the pink surface.

\section*{Experiment}
We experimentally investigate the sudden transition phenomenon utilizing a single NV center in diamond. 
Figure \ref{Fig2}(a) shows the two-qutrit system constructed by the electron spin of the NV center and the nuclear spin of the $^{14}{\rm N}$ atom both with spin-1.
With a static magnetic field applied along the NV symmetry axis, the Hamiltonian of this two-qutrit spin system can be written as
\begin{align}
	H_{\rm NV} = 2\pi(DS_z^2+\omega_{\rm e}S_z+PI_z^2+\omega_{ \rm n}I_z+A_{\rm hf}S_zI_z),
\end{align}
where $S_z\ (I_z) $ is the spin-1 operator of the electron (nuclear) spin, $D = 2.87$ GHz is the electronic zero-field splitting, $P = -4.95$ MHz represents the nuclear quadrupolar interaction, and $A_{\rm hf} = -2.16$ MHz is the hyperfine coupling constant.
$\omega_{\rm e}$ ($\omega_{\rm n}$) corresponds to the Zeeman frequency of the electron (nuclear) spin.
The energy level structure of the two-qutrit system is depicted in Fig. \ref{Fig2}(a) right.
The eigenlevels are denoted by $|m_S\rangle_{\rm e}\otimes|m_I\rangle_{\rm n}$, with $m_S,m_I=0,\,\pm1$ representing the states of the electron and nuclear spins, respectively.
The state of the electron spin can be manipulated by microwave (MW) pulses, which are labeled by the purple arrows in Fig. \ref{Fig2}(a).
The radio frequency (RF) pulses labeled by blue arrows were performed to control the state of the nuclear spin.
For simplicity, $|{+1}\rangle_{\rm e(n)}$, $|0\rangle_{\rm e(n)}$, and $|{-1
}\rangle_{\rm e(n)}$ are hereafter labeled by $|0\rangle$, $|1\rangle$, and $|2\rangle$ , respectively. And $|m_S\rangle_{\rm e}\otimes |m_I\rangle_{\rm n}$ is labeled as the corresponding $|ij\rangle $ with $\ i,j = 0,1,2$.

Relaxation processes that cause the decoherence and energy dissipation of the NV center system can be characterized by the dephasing time $T_{2,\rm e(n)}^* $ and the longitudinal relaxation time $T_{1,\rm e(n)}$, respectively.
In our experiment, the diamond sample was isotopically purified with the concentration of $^{12}{\rm C}$ atom exceeding $99.9\%$.
For the electron spin, the dephasing time is measured to be $T_{\rm 2,e}^* = 44 \pm 2\ {\rm \mu}$s on average (See supplemental material).
The longitudinal relaxation times of the electron spin and the nuclear spin, the dephasing time of the nuclear spin are all over one millisecond \cite{PRL_2022_Fu}.
For the timescale concerned in this experiment ($<100$ $\mu$s), the influence of these noises is limited. We omit them hereafter and only consider the dephasing noise of the electron spin.
Therefore, the two-qutrit spin system can be taken as governed by the one-qutrit local dephasing noise and follows the evolution in Eq. \ref{Dephased_BD_States} after the appropriate initial state was prepared.

Figure \ref{Fig2}(b) shows the pulse sequence diagram for studying the dynamical behaviors of various QCs.
It consists of four parts: polarization, state preparation, evolution under the noise environment, and measurement.
The NV center was polarized into state $|10\rangle$ via a green laser pulse with the external magnetic field being 500 Gauss \cite{PRL_2009_Jacques}.
To prepare the two-qutrit Bell-diagonal state, a MW pulse was performed to convert the polarized state $|10\rangle $ to $\sqrt{c_0}|10\rangle +\sqrt{c_2}|20\rangle$.
After a waiting time of $t_{\rm wait} = 200 \ \mu$s, the system state evolved into the mixed state $\rho_{\rm I}=c_0|10\rangle\langle10| +c_2|20\rangle\langle20|$ due to the dephasing.
Then four selective RF pulses followed by the same waiting time were executed to prepare the NV center into state $\rho_{\rm II}=\frac{c_0}{3}\left(|10\rangle+|11\rangle+|12\rangle\right)(\langle10|+\langle11|+\langle12|)+\frac{c_2}{3}(|20\rangle+|21\rangle+|22\rangle)(\langle20|+\langle21|+\langle22|)
$. 
Finally, selectively MW pulses were applied sequentially to eventually realize the Bell-diagonal state $\rho_{\rm BD,3}$ (see Fig. \ref{Fig2}(b)).
The selective pulses mentioned above correspond to transitions between different energy levels displayed in Fig.~\ref{Fig2}(a).
The state $\rho_{\rm BD,3}$ with different parameters can be prepared by adjusting the time duration of the pulses.
After the state was prepared, the quantum system was left to evolve under the dephasing noise for a period of $t$.
In the final part, a set of pulse sequences (dashed box in Fig. \ref{Fig2}(b)) were executed to reconstruct the quantum state.
More details about the state preparation, the measurement, and the state reconstruction can be found in the supplementary information.

\begin{figure}\centering
	\includegraphics[width=0.8\columnwidth]{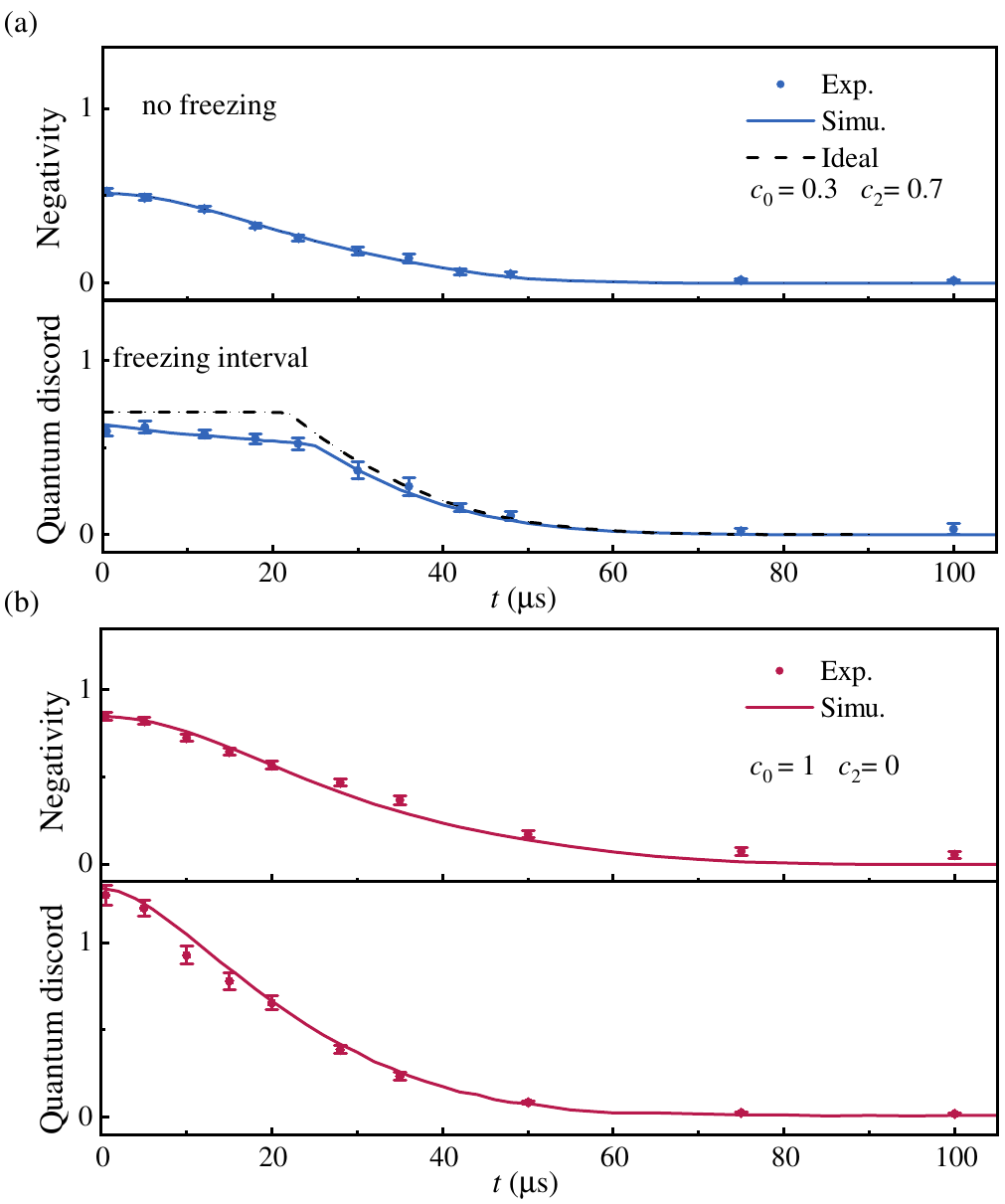}
	\caption{Experimental dynamics of high-dimensional quantum entanglement and quantum discord.
		The x-axis is the evolution time under one-qutrit local dephasing noise.
		The y-axis is the quantity of entanglement (characterized by negativity) or discord.
		The dots with error bars are experimental data, the lines show the simulation results.
		(a) Parameters $c_0=0.3, c_2=0.7$. The negativity decayed smoothly while the discord dynamics presented a sudden transition.
		{ The black dashed line presents the ideal freezing dynamics of the discord without the influence of imperfect polarization.}  
		(b) Parameters $c_0=1, c_2=0$.
		Both the discord and entanglement decay smoothly to 0.
	}
	\label{Fig3}
\end{figure}

The quantity of various QCs was obtained from the measured density matrices of the two-qutrit system.
To ascertain the physical validity of the matrices, maximum likelihood estimation was applied to the experimental data \cite{PRA_2001_James}.
Due to imperfect polarization and statistical errors, the obtained density matrices may deviate slightly from the form shown in Eq. \ref{Dephased_BD_States}.
To accurately calculate the discord of these quantum states, we adopted the quantum mutual information description \cite{PRL_2001_Olivier} and utilized a numerical method \cite{PRA_2012_Rossignoli}.
We also explored the dynamics of quantum entanglement under the same local dephasing noise.
It is quantified by negativity defined as $\mathcal{N} = ({||} \rho^{PT}{||}_1 -1)/2$ with $PT$ representing partial transposition \cite{PRA_2007_Derkacz}.
Details about the quantum mutual information description and calculation of different kinds of QCs are included in the supplementary information.

\section*{Results}
Figure \ref{Fig3} (a) shows the experimental dynamics of high-dimensional quantum discord and quantum entanglement with parameters $c_0 = 0.3, c_2 = 0.7$.
The dots with error bars are experimental data while the solid lines are simulation results.
In the top panel, the entanglement decays smoothly as the evolution time $t$ increases.
In contrast, the discord is almost a constant for $t\lesssim 23\ \mu s$, which is consistent with the frozen part in Fig. \ref{Fig1} and shows the immunity of discord against the dephasing noise.
When $t\gtrsim 23\ \mu s$, the discord decays gradually to 0.
Therefore, the dynamics of the discord presented in the bottom panel agree well with the theoretical anticipation, and the freezing phenomenon was observed experimentally.
It is noticed that the experimental discord dynamics show a slight decay and deviate from the ideal situation (the black dashed line) during the freezing interval. 
This is due to the imperfect polarization of the spins \cite{PRB_2013_Fischer, PRL_2009_Jacques}. 
In our experiment, the polarization rate of the electron spin and the nuclear spin were 0.92(1) and 0.98(1), respectively.
Corresponding simulation results are plotted by the solid blue line and agree well with the experiment.

The dynamics of entanglement and discord when $c_0 = 1, c_2 = 0$ were also investigated.
The evolution trajectory of the Bell-diagonal state with these parameters corresponds to the red line in Fig. \ref{Fig1}(b).
This trajectory does not cross the pink surface where $d_1=d_2$, so the sudden transition shall not happen. 
The experimental results are depicted in Fig. \ref{Fig3} (b).
The dynamics of both discord and entanglement exhibit a gradual decay as the evolution time, $t$, progresses. 
This behavior is in line with our theoretical predictions.

After experimental demonstrating the robustness of discord against the dephasing noise in qutrit system, we now compare it with its qubit counterpart.
It has shown that there is a certain trade-off between the amount of discord and the time that it can be frozen \cite{review_2018_Bera}.
This trade-off is vital when discord is utilized in QIP as the speedup of QIP tasks depends severely on the quantity of quantum correlations \cite{Review_2017_Cham}.
To quantify the goodness of the freezing behavior, the concept of freezing index was introduced \cite{PRA_2015_Chanda}. 
When the freezing phenomenon starts from the beginning of the dynamics and occurs solely once, the freezing index can be defined as:
\begin{equation}
    \mathcal{F}=\left\{~\overline{\mathcal{Q}}\int_{\gamma_{\rm ini}}^{\gamma_{\rm fin}}\mathcal{Q}(\gamma)d\gamma~\right\}^{1/4}
\end{equation}
where $\gamma:=1-\lambda$ is the parameterized time and $\gamma_{\rm fin}-\gamma_{\rm ini}$ denotes the freezing interval. $\mathcal{Q}(\gamma)$ ($\overline{\mathcal{Q}}$) is the time-dependent (averaged) discord during the freezing interval when imperfect freezing was considered and the discord decays slowly during the period.

\begin{figure}\centering
\includegraphics[width=1\columnwidth]{
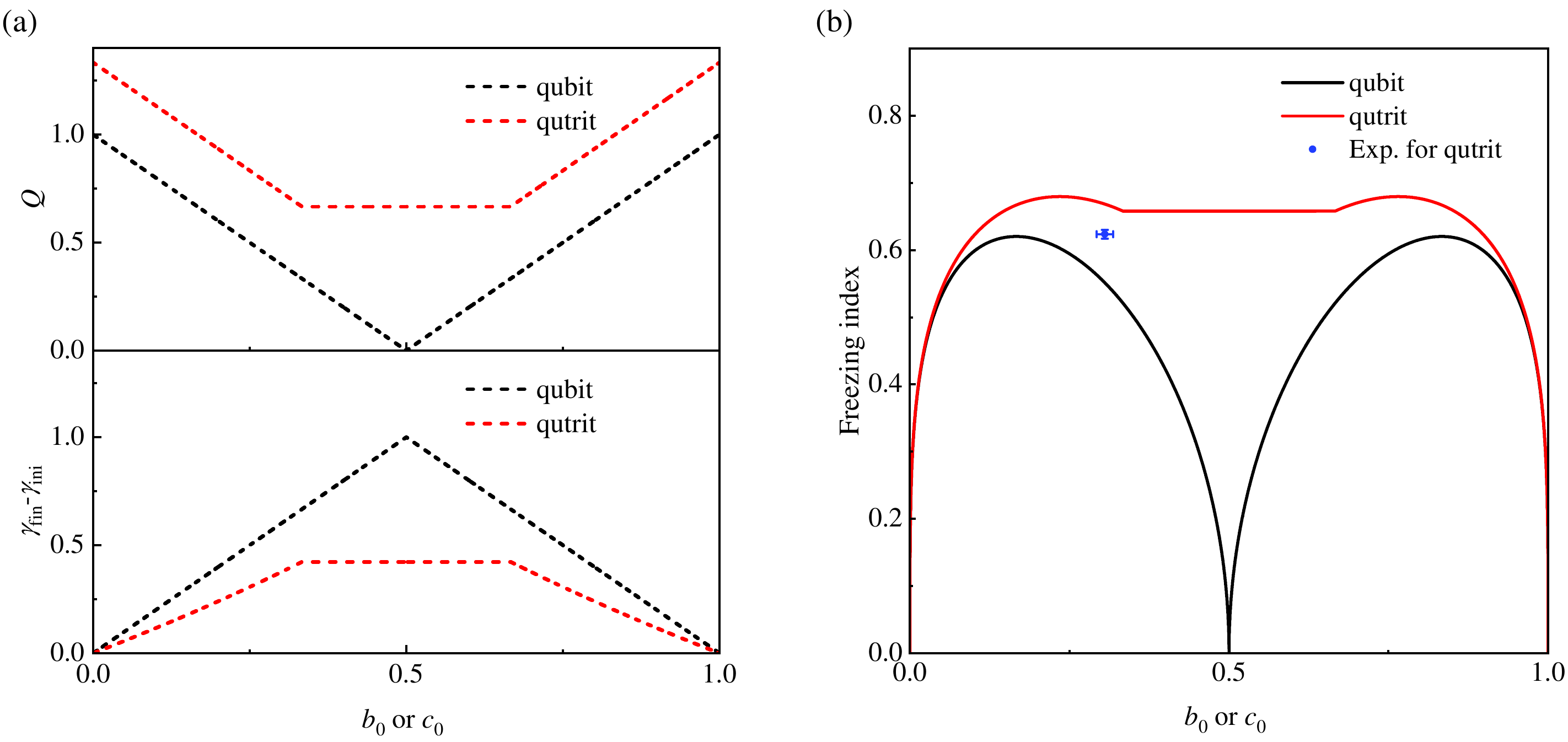}
\caption{\textbf{Freezing index comparison between qubit and qutrit.} (a) The amount of discord (upper part) and the freezing interval (lower part) versus qubit parameter $b_0$ or qutrit parameter $c_0$. (b) Freezing index versus $b_0$ and $c_0$. The dot with error bars is the experimental result.}
\label{Fig4}
\end{figure}

We compare the freezing index of the two-qutrit system with the two-qubit system for Bell-diagonal states.
The qubit Bell-diagonal states are defined as $\rho_{\rm BD,2} =b_0|\Phi_0\rangle \langle\Phi_0|+b_1|\Phi_1\rangle \langle\Phi_1|$ where $|\Phi_0\rangle$ and $|\Phi_1\rangle$ are qubit maximally states, $b_0, b_1\in [0,1]$ with $b_0+b_1 = 1$.
Fig. \ref{Fig4}(a) illustrates the trade-off between the discord value and the freezing interval.
It can be seen that a longer freezing interval usually accompanies a smaller amount of discord for both qubits and qutrits.
And while the freezing interval of qubits is always longer than qutrits, the discord contained in qubit states is smaller than that in qutrits.
This trade-off is finally characterized by the freezing index given in Fig. \ref{Fig4}(b).
It is clear that the freezing index of qutrits exceeds that for qubits consistently, showing qutrit Bell-diagonal states are more potential than those of qubit for QIP when suffering local dephasing noise. 
The dot with error bars is the experimental result, whose freezing index is bigger than the theoretical value of qubit states with the same parameter $b_0=c_0=0.3$.
The distance between the experimental data and the theoretical value of the qutrit freezing index comes mainly from imperfect polarization and other statistical errors.	

\section*{Conclusion}
We have conducted an in-depth investigation into the dynamics of high-dimensional QCs under the influence of local dephasing noise.
A freezing phenomenon in the high-dimensional quantum discord dynamics was observed, which is consistent with the prediction of the geometric picture we proposed, while QE decays monotonously.
Comparison between this phenomenon and that in qubits shows qutrit Bell-diagonal states have a larger freezing index.
Our work demonstrates the potential of utilizing high-dimensional quantum discord to implement QIP tasks as dephasing noises are common in most quantum systems.
Besides, the behavior of high-dimensional quantum discord under other noises could be explored. 
For example, whether quantum discord exhibits a sudden death under depolarization noise would be an interesting issue.

We thank Zhu-Jun Zheng and Changyue Zhang for the helpful discussion.
This work was supported by the National Natural Science Foundation of China (Grants No. 12261160569, No. 12321004, and No. 12374462), the Chinese Academy of Sciences (Grants No. XDC07000000 and No. GJJSTD20200001), Innovation Program for Quantum Science and Technology (Grant No. 2021ZD0302200), Anhui Initiative in Quantum Information Technologies (Grant No. AHY050000), and Hefei Comprehensive National Science Center, the Fundamental Research Funds for the Central Universities (Grants No. 226-2023-00137 and No. 226-2023-00139).
Y.F. acknowledges financial support from the China Postdoctoral Science Foundation (Nos. 2023TQ0029, 2023M740262
and GZC20233415).

\onecolumngrid
\vspace{1.5cm}
\begin{center}
	\newpage
	\textbf{\large Supplementary Material}
\end{center}

\setcounter{figure}{0}
\setcounter{equation}{0}
\setcounter{table}{0}
\makeatletter
\renewcommand{\thefigure}{S\arabic{figure}}
\renewcommand{\theequation}{S\arabic{equation}}
\renewcommand{\thetable}{S\arabic{table}}
\renewcommand{\bibnumfmt}[1]{[RefS#1]}
\renewcommand{\citenumfont}[1]{RefS#1}

\renewcommand{\figurename}{Supplementary Figure}
\renewcommand{\tablename}{Supplementary Table}
\section{Zero-Discord States in the Parameter Space}

In the main text, we elucidated that the ensemble of zero-discord states within the parameter space is bifurcated, encompassing the line $\chi_1$ and the plane $\chi_2$.
Herein, we offer a meticulous exposition. 
The evolution of the two-qutrit Bell-diagonal states, subject to the local dephasing noise of the first qutrit, is
\begin{small}
\begin{align}
&\rho_{\rm BD}^{\rm deph}(c_0,c_2,\lambda)\notag\\
=	&\frac{1}{3}\left(\begin{array}{ccccccccc}
		c_0&  0&  0&  0&c_0\lambda&  0&  0&  0&c_0  \lambda^2\\
		0&c_1&   0&    0&    0&c_1\lambda&c_1\lambda^2&    0&   0\\
		0&    0&c_2&c_2\lambda&   0&    0&    0&c_2\lambda^2&   0\\
		0&    0&c_2\lambda&c_2&    0&    0&    0&c_2\lambda&   0\\
		c_0\lambda&    0&    0&    0&c_0&    0&    0&    0&c_0\lambda  \\
		0&c_1\lambda&    0&    0&    0&c_1&c_1\lambda&    0&   0\\
		0&c_1\lambda^2&    0&    0&    0&c_1\lambda&c_1&    0&   0\\
		0&    0&c_2\lambda^2&c_2\lambda&    0&    0&    0&c_2&   0\\
		c_0\lambda^2&    0&    0&    0&c_0\lambda&    0&    0&    0&c_0  
	\end{array}\right),
	\label{eq4}
\end{align}
\end{small}
where $c_1=1-c_0-c_2$.
The dynamics are characterized by three variables: $c_0$, $c_2$, and $\lambda$, which collectively constitute a triangular prism-shaped parameter space, as illustrated in Figure 1 of the main text. 
The necessary and sufficient criterion for the quantum discord (QD) of a biparticle quantum state $\rho_{AB}$ to vanish is that $\rho_{AB}$ belongs to quantum-classical states \cite{PRL_2010_Dakić} with the form
\begin{equation}
\rho_{\rm q-c}=\sum_{i}p_i \rho_i\otimes |\phi_i\rangle\langle \phi_i| .
\end{equation}
Here $p_i\ge0$ represents the probability associated with each state $|\phi_i\rangle$ and satisfy $\sum_ip_i=1$, $\langle \phi_i|\phi_j\rangle = \delta_{ij}$ ensures that the states $|\phi_i\rangle$ are mutually exclusive, and $\rho_i$ denotes a density matrix of subsystem $A$.
A crucial property of quantum-classical states is that there exists a von Neumann measurement on subsystem $B$, which leaves the state of subsystem $A$ undisturbed.
Therefore, the QD of quantum-classical states is 0.

\begin{figure}
\centering
\includegraphics[width=0.9\columnwidth]{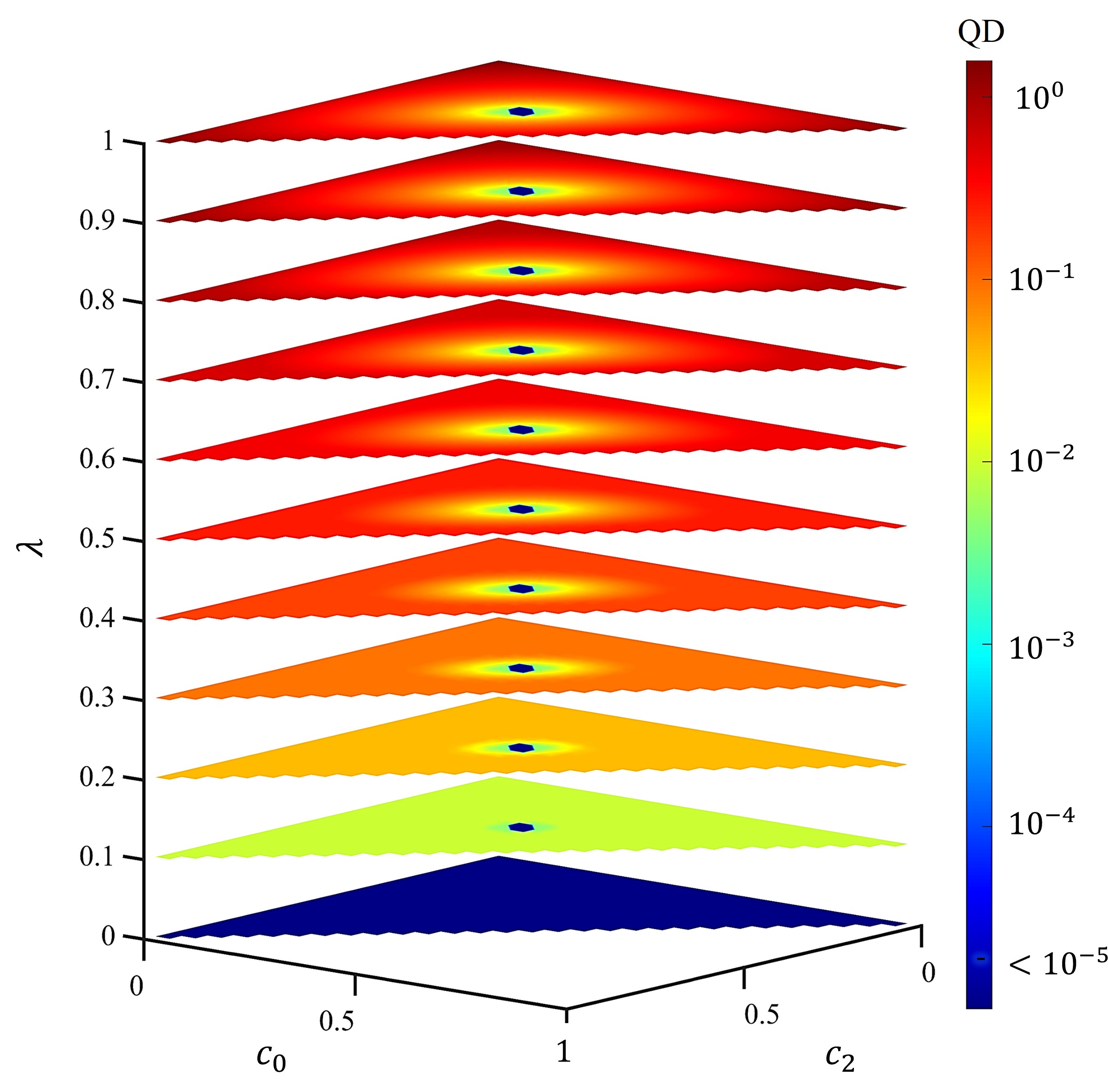}
\caption{\textbf{Numerical Results of Quantum Discord in The Parameter Space.}
The numerical results suggest that the quantum-classical states with zero QD contain two parts: the line $\chi_1$ where $c_0=c_2=1/3$ and the plane $\chi_2$ where $\lambda=0$.
}
\label{zero_QD_States}
\end{figure}

The states in $\chi_1$ satisfy $c_0=c_2=1/3$ and can be written as 
\begin{equation}
    \rho_{\chi_1}=	\frac{1}{9}\left(
    \begin{array}{ccccccccc}
	1	&  0&  0&  0&\lambda&  0&  0&  0&  \lambda^2\\
			0&1&   0&    0&    0&\lambda&\lambda^2&    0&   0\\
			0&    0&1&\lambda&   0&    0&    0&\lambda^2&   0\\
			0&    0&\lambda&1&    0&    0&    0&\lambda&   0\\
			\lambda&    0&    0&    0&1&    0&    0&    0&\lambda  \\
			0&\lambda&    0&    0&    0&1&\lambda&    0&   0\\
			0&\lambda^2&    0&    0&    0&\lambda&1&    0&   0\\
			0&    0&\lambda^2&\lambda&    0&    0&    0&1&   0\\
			\lambda^2&    0&    0&    0&\lambda&    0&    0&    0&1  
		\end{array}
  \right).
\end{equation}
They can be decomposed into $\rho_{\chi_1}=\frac{1}{3}\left( \rho_0|\phi_0\rangle \langle \phi_0|+\rho_1|\phi_1\rangle \langle \phi_1|+\rho_2|\phi_2\rangle \langle \phi_2|\right)$ with
\begin{equation}
    \begin{array}{l}
   |\phi_0\rangle=\frac{1}{\sqrt{3}} \left( \begin{array}{c} 1\\ 1\\ 1 \end{array} \right),
   \rho_0= \frac{1}{3}
		\left(\begin{array}{ccc}
		1& \lambda& \lambda^2\\
		\lambda& 1& \lambda\\
		\lambda^2&  \lambda& 1
		\end{array}  
		\right),\\
   |\phi_1\rangle= \frac{1}{\sqrt{3}}\left( \begin{array}{c} \frac{-1+\sqrt{3}\rm i}{2}\\ 1\\ \frac{-1-\sqrt{3}\rm i}{2} \end{array} \right),\\
   |\phi_2\rangle=\frac{1}{\sqrt{3}} \left( \begin{array}{c} \frac{-1-\sqrt{3}\rm i}{2}\\ 1\\ \frac{-1+\sqrt{3}\rm i}{2} \end{array} \right),\\
   \rho_1=\frac{1}{3}\left(\begin{array}{ccc}
			1& \frac{-1-\sqrt{3}\rm i}{2}\lambda&\  \frac{-1+ \sqrt{3}\rm i}{2}\lambda^2\\
			\frac{-1+ \sqrt{3}\rm i}{2}\lambda& 1&\ \frac{-1- \sqrt{3}\rm i}{2}\lambda\\
			\frac{-1- \sqrt{3}\rm i}{2}\lambda^2&  \frac{-1+ \sqrt{3}\rm i}{2}\lambda& 1
		\end{array}
		\right),\\
   
   \rho_2=\frac{1}{3}\left(\begin{array}{ccc}
				1& \frac{-1+\sqrt{3}\rm i}{2}\lambda&\  \frac{-1- \sqrt{3}\rm i}{2}\lambda^2\\
				\frac{-1- \sqrt{3}\rm i}{2}\lambda& 1&\ \frac{-1+ \sqrt{3}\rm i}{2}\lambda\\
				\frac{-1+ \sqrt{3}\rm i}{2}\lambda^2&  \frac{-1- \sqrt{3}\rm i}{2}\lambda& 1
			\end{array}
			\right).
    \end{array}
\end{equation}

The states in $\chi_2$ satisfy $\lambda=0$ and can be decomposed into 
\begin{align}
    \rho_{\chi_2}&=	\frac{1}{3}\left(\begin{array}{ccccccccc}
		c_0&  0&  0&  0&0&  0&  0&  0&0\\
		0&c_1&   0&    0&    0&0&0&    0&   0\\
		0&    0&c_2&0&   0&    0&    0&0&   0\\
		0&    0&0&c_2&    0&    0&    0&0&   0\\
		0&    0&    0&    0&c_0&    0&    0&    0&0  \\
		0&0&    0&    0&    0&c_1&0&    0&   0\\
		0&0&    0&    0&    0&0&c_1&    0&   0\\
		0&    0&0&0&    0&    0&    0&c_2&   0\\
		0&    0&    0&    0&0&    0&    0&    0&c_0  
	\end{array}\right)\notag \\
 &= \frac{1}{3}\left( \rho_0'|\phi_0'\rangle \langle \phi_0'|+\rho_1'|\phi_1'\rangle \langle \phi_1'|+\rho_2'|\phi_2'\rangle \langle \phi_2'|\right).
\end{align}
with
\begin{equation}
    \begin{array}{cc}
   |\phi_0^\prime\rangle= \left( \begin{array}{c} 1\\ 0\\ 0 \end{array} \right),  &
    \rho_0^\prime= 
		\left(\begin{array}{ccc}
		c_0 &  0  &  0    \\
		0   & c_2 &  0    \\
		0   &  0  & c_1
		\end{array}  
		\right),\\
   |\phi_1^\prime\rangle= \left( \begin{array}{c} 0\\ 1\\ 0 \end{array} \right),  &
   \rho_1^\prime=
       \left(\begin{array}{ccc}
		c_1 &  0  &  0    \\
		0   & c_0 &  0    \\
		0   &  0  & c_2
		\end{array}
		\right),\\
   |\phi_2^\prime\rangle= \left( \begin{array}{c} 0\\ 0\\ 1 \end{array} \right), &
   \rho_2^\prime=
       \left(\begin{array}{ccc}
		c_2 &  0  &  0    \\
		0   & c_1 &  0    \\
		0   &  0  & c_0
		\end{array}
			\right).   
    \end{array}
\end{equation}
Hence, any quantum state residing within either $\chi_1$ or $\chi_2$ is categorized as a quantum-classical state, thereby possessing a QD value of zero. 
Regarding other quantum states, analytically demonstrating their non-zero QD remains a challenging task. 
Consequently, we resorted to numerical methods, and the outcomes presented in Fig. \ref{zero_QD_States} suggest that these states indeed exhibit a non-zero QD.

\section{Equation of the $d_1=d_2$ Surface and the Sudden Transition}

In this section, we derive the equation governing the surface $d_1=d_2$ depicted in Fig. \ref{Fig1} of the main text and demonstrate the occurrence of the sudden transition phenomenon.

Consider a Bell-diagonal state $\rho_{\rm P}=\rho_{\rm BD}^{\rm deph}(c_0,c_2,\lambda)$, which represents a specific point P within the parameter space. 
The QD of this state is determined by the minimum of $d_1$ and $d_2$. 
Here, $d_1$ represents the minimum Schatten 1-norm trace distance between $\rho_{\rm P}$ and any quantum state belonging to $\chi_1$, while $d_2$ signifies the corresponding minimum distance to quantum states in $\chi_2$. 
We posit that the quantum state within $\chi_1$ (or $\chi_2$) that is closest to $\rho_{\rm P}$ is identified by the intersection point of the perpendicular line extending from point P to the line $\chi_1$ (or plane $\chi_2$). 
The detailed derivations underlying these assertions are outlined below.

\begin{itemize}
\item Closest quantum state to $\rho_{\rm P}$ in the line $\chi_1$.  \\
A quantum state $\sigma_1$ in $\chi_1$ has the form $\sigma_1(\lambda')=\rho_{\rm BD}^{\rm deph}(1/3,1/3,\lambda^\prime)$.
The Schatten 1-norm trace of it to $\rho_{\rm P}$ can be calculated as 

\begin{footnotesize}
\begin{equation}
\begin{aligned}
    &~~~||\rho_{\rm P}-\sigma_1(\lambda')||_1 = ||\rho_{\rm BD}^{\rm deph}(c_0,c_2,\lambda)-\rho_{\rm BD}^{\rm deph}(1/3,1/3,\lambda^\prime)||_1   \\
    &=\left|\left| \frac{1}{9}\left(\begin{array}{ccccccccc}
	  3c_0-1&        0&        0&        0&    3c_0\lambda-\lambda^\prime&           0&          0&          0&   3c_0\lambda^2-{\lambda^\prime}^2\\
	 	0&    3c_1-1&        0&        0&          0&     3c_1\lambda-\lambda^\prime&  3c_1\lambda^2-{\lambda^\prime}^2&          0&           0\\
		0&        0&      3c_2-1&  3c_2\lambda-\lambda^\prime&          0&           0&          0& 3 c_2\lambda^2-{\lambda^\prime}^2&           0\\
		0&        0&  3c_2\lambda-\lambda^\prime&      3c_2-1&          0&           0&          0&    3c_2\lambda-\lambda^\prime&           0\\
      3c_0\lambda-\lambda^\prime&        0&        0&        0&        3c_0-1&           0&          0&          0&     3c_0\lambda-\lambda^\prime\\
		0&  3c_1\lambda-\lambda^\prime&        0&        0&          0&         3c_1-1&     3c_1\lambda-\lambda^\prime&          0&           0\\
		0&3c_1\lambda^2-{\lambda^\prime}^2&        0&        0&          0&     3c_1\lambda-\lambda^\prime&       3c_1-1&          0&           0\\
		0&        0&3c_2\lambda^2-{\lambda^\prime}^2&  3c_2\lambda-\lambda^\prime&          0&           0&          0&        3c_2-1&           0\\
    3c_0\lambda^2-{\lambda^\prime}^2&        0&        0&        0&    3c_0\lambda-\lambda^\prime&           0&          0&          0&         3c_0-1 
	\end{array}\right) \right|\right|_1  \\
    &=\left|\left| \frac{1}{9}\left(\begin{array}{ccccccccc}
	                3c_0-1&    3c_0\lambda-\lambda^\prime&    3c_0\lambda^2-{\lambda^\prime}^2&    0&    0&     0&    0&    0&   0\\
        3c_0\lambda-\lambda^\prime&                    3c_0-1&          3c_0\lambda-\lambda^\prime&    0&    0&     0&    0&    0&   0\\
  3c_0\lambda^2-{\lambda^\prime}^2&    3c_0\lambda-\lambda^\prime&                          3c_0-1&    0&    0&     0&    0&    0&   0\\
        0&    0&    0&                        3c_1-1&   3c_1\lambda-\lambda^\prime&   3c_1\lambda^2-{\lambda^\prime}^2&   0&    0&   0\\
        0&    0&    0&        3c_1\lambda-\lambda^\prime&                   3c_1-1&         3c_1\lambda-\lambda^\prime&   0&    0&   0\\
	0&    0&    0& 3c_1 \lambda^2-{\lambda^\prime}^2&  3c_1 \lambda-\lambda^\prime&                         3c_1-1&   0&    0&   0\\
	0&    0&    0&    0&   0&   0&                        3c_2-1&   3c_2\lambda-\lambda^\prime&   3c_2\lambda^2-{\lambda^\prime}^2\\
	0&    0&    0&    0&   0&   0&       3c_2 \lambda-\lambda^\prime&                   3c_2-1&         3c_2\lambda-\lambda^\prime\\
        0&    0&    0&    0&   0&   0&  3c_2\lambda^2-{\lambda^\prime}^2&   3c_2\lambda-\lambda^\prime&                         3c_2-1 
	\end{array}\right) \right|\right|_1     \\
    &=||\tilde{\rho}(c_0,\lambda,\lambda^\prime)||_1+||\tilde{\rho}(c_1,\lambda,\lambda^\prime)||_1+||\tilde{\rho}(c_2,\lambda,\lambda^\prime)||_1 ,
\label{eq7}
\end{aligned}
\end{equation}    
\end{footnotesize}

where 
\begin{small}
\begin{align}
&\tilde{\rho}(c_i,\lambda,\lambda^\prime)\notag\\
= &\frac{1}{9}\left(\begin{array}{ccc}
	                      3c_i-1&    3c_i\lambda-\lambda^\prime&    3c_i\lambda^2-{\lambda^\prime}^2\\
       3c_i\lambda-\lambda^\prime&                         3c_i-1&          3c_i\lambda-\lambda^\prime\\
 3c_i\lambda^2-{\lambda^\prime}^2&     3c_i\lambda-\lambda^\prime&                              3c_i-1
	\end{array}\right) .
\end{align}
\end{small}
Let us define the three eigenvalues of $\tilde{\rho}(c_i,\lambda,\lambda^\prime)$ as $\epsilon_{i,1}$, $\epsilon_{i,2}$, and $\epsilon_{i,3}$. 
Then, it follows that $\epsilon_{i,1}+\epsilon_{i,2}+\epsilon_{i,3}={\rm Tr}[\tilde{\rho}(c_i,\lambda,\lambda^\prime)]=c_i-1/3$.
Additionally, $||\tilde{\rho}(c_i,\lambda,\lambda^\prime)||_1 = \sum_{j=1}^3|\epsilon_{i,j}|\ge|\sum_{j=1}^3\epsilon_{i,j}|=|c_i-1/3|:=d_{\rm min,1}$.
The quantum state that corresponds to the intersection point of the perpendicular line extending from point P to the line $\chi_1$ is denoted as
$\sigma_1(\lambda)$. 
The Schatten 1-norm trace distance between $\sigma_1(\lambda)$ and $\rho_{\rm P}$ is given by $||\rho_{\rm P}-\sigma_1(\lambda)||_1=\sum_{i=0}^2||\tilde{\rho}(c_i,\lambda,\lambda)||_1$.
The three eigenvalues of $\tilde{\rho}(c_i,\lambda,\lambda)$ are $\tilde{\epsilon}_{i,1}=(3c_i-1)(\lambda^2-1)/9$ and $\tilde{\epsilon}_{i,2|3}=[(3c_i-1)(\lambda^2+2)\pm\sqrt{(3c_i-1)^2\lambda^2(8+\lambda^2)}]/18$.
It is noteworthy that $|(3c_i-1)(\lambda^2+2)|\ge|\sqrt{(3c_i-1)^2\lambda^2(8+\lambda^2)}|$.
Consequently, $||\tilde{\rho}(c_i,\lambda,\lambda)||_1=|\tilde{\epsilon}_{i,1}|+|\tilde{\epsilon}_{i,2}|+|\tilde{\epsilon}_{i3}|=|c_i-1/3|=d_{\rm min,1}$.
This establishes that the minimum value of $||\tilde{\rho}(c_i,\lambda,\lambda^\prime)||_1$, as well as $||\rho_{\rm P}-\sigma_1(\lambda)||_1$, is attained when $\lambda^\prime=\lambda$.
Therefore, we have rigorously demonstrated that the quantum state within the set $\chi_1$ that is closest to $\rho_{\rm P}$ corresponds precisely to the intersection point of the perpendicular line extending from point P to the line $\chi_1$.

\item Closest quantum state to $\rho$ in the plane $\chi_2$.

A quantum state $\sigma_2$ in $\chi_2$ has the form $\sigma_2(c_0^\prime,c_2^\prime)=\rho_{\rm BD}^{\rm deph}(c_0',c_2',0)$.
Similar to the transformation of Eq.(\ref{eq7}), the Schatten 1-norm trace of $\sigma_2(c_0^\prime,c_2^\prime)$ to $\rho_{\rm P}$ can be calculated as 

\begin{equation}
\begin{aligned}
    &~~~~||\rho_{\rm P}-\sigma_2(c_0^\prime,c_2^\prime)||_1 = ||\rho_{\rm BD}^{\rm deph}(c_0,c_2,\lambda)-\rho_{\rm BD}^{\rm deph}(c_0',c_2',0)||_1   \\
    &=\left|\left| \frac{1}{3}\left(\begin{array}{ccccccccc}
	   c_0-c_0'&    c_0\lambda&    c_0\lambda^2&    0&    0&     0&    0&    0&   0\\
        c_0\lambda&      c_0-c_0'&      c_0\lambda&    0&    0&     0&    0&    0&   0\\
      c_0\lambda^2&    c_0\lambda&        c_0-c_0'&    0&    0&     0&    0&    0&   0\\
        0&    0&    0&      c_1-c_1'&   c_1\lambda&   c_1\lambda^2&   0&    0&   0\\
        0&    0&    0&    c_1\lambda&     c_1-c_1'&     c_1\lambda&   0&    0&   0\\
	0&    0&    0& c_1 \lambda^2&   c_1\lambda&       c_1-c_1'&   0&    0&   0\\
	0&    0&    0&    0&   0&   0&      c_2-c_2'&   c_2\lambda&   c_2\lambda^2\\
	0&    0&    0&    0&   0&   0&    c_2\lambda&     c_2-c_2'&      c_2\lambda\\
        0&    0&    0&    0&   0&   0&  c_2\lambda^2&   c_2\lambda&         c_2-c_2' 
	\end{array}\right) \right|\right|_1     \\
    &=||\tilde{\rho}(c_0,c_0',\lambda)||_1+||\tilde{\rho}(c_1,c_1',\lambda)||_1+||\tilde{\rho}(c_2,c_2',\lambda)||_1 ,
\label{eq9}
\end{aligned}
\end{equation}    

where 
\begin{equation}
\tilde{\rho}(c_i,c_i^\prime,\lambda)= \frac{1}{3}\left(\begin{array}{ccc}
    c_i-c'_i&   c_i\lambda&    c_i\lambda^2\\
  c_i\lambda&     c_i-c'_i&      c_i\lambda\\
c_i\lambda^2&   c_i\lambda&        c_i-c'_i
	\end{array}\right) .
\end{equation}

The three eigenvalues of $\tilde{\rho}(c_i,c'_i,\lambda)$ are $\nu_{i,1} = c_i-c'_i-c_i\lambda^2$, and $\nu_{i,2|3} = c_i-c'_i+\frac{1}{2}c_i(\lambda^2\pm\sqrt{\lambda^4+8\lambda^2})$. 
Accordingly,
\begin{align}
||\rho_{\rm P}-\sigma_2(c_0^\prime,c_2^\prime)||_1
&=\sum_{i=0}^{2}\left(|\nu_{i,1}|+|\nu_{i,2}|+|\nu_{i,3}| \right)
\ge|\sum_{i=0}^{2}\nu_{i,1}|+\sum_{i=0}^{2}|(\nu_{i,2}-\nu_{i,3})|   \\
&=|\sum_{i=0}^{2}c_i-\sum_{i=0}^{2}c_i'-\sum_{i=0}^{2}c_i\lambda^2|+\sum_{i=0}^{2}|c_i\sqrt{\lambda^4+8\lambda^2}|\\
&=\lambda^2+\sqrt{\lambda^4+8\lambda^2} :=d_{\rm min,2}.
\end{align}

It is noticed that when $c_0=c_0'$ and $ c_2 =c_2'$
\begin{align}
  &||\rho_{\rm P}-\sigma_2(c_0,c_2)||_1\notag\\
  =& \lambda^2+\frac{1}{2}\left|\lambda^2-\sqrt{\lambda^4+8\lambda^2}\right|+ \frac{1}{2}\left|\lambda^2+\sqrt{\lambda^4+8\lambda^2}\right|\notag \\
  =&\lambda^2+\sqrt{\lambda^4+8\lambda^2} =d_{\rm min,2}
\end{align}

Therefore, we have proved that for any quantum states $\sigma_2\in\chi_2$, the minimum value of $||\rho_{\rm P}-\sigma_2(c_0^\prime,c_2^\prime)||_1$ is achieved when $c_0 = c_0',c_2 = c_2'$.
This minimum value corresponds precisely to the quantum state that lies at the intersection point of the perpendicular line extending from point P to the plane $\chi_2$.

\end{itemize}

To sum up, we have calculated and demonstrated that $d_1=\sum_{i=0}^3|c_i-1/3|$ and $d_2=\lambda^2+\sqrt{\lambda^4+8\lambda^2}$. 
Equation of the $d_1=d_2$ surface can be formulated as $\lambda\sqrt{\lambda^2+8}+\lambda^2 = |1-3c_0|+|1-3c_1|+|1-3c_2|$.
For any quantum state within the parameter space, its QD is ${\rm min}\{\sum_{i=0}^3|c_i-1/3|,~\lambda^2+\sqrt{\lambda^4+8\lambda^2}\}$.

It is noteworthy that $d_1$ does not depend on $\lambda$ and $d_2$ is independent of both $c_0$ and $c_2$. 
The trajectory traced by $\rho_{\rm BD}^{\rm deph}$ within the parameter space corresponds to a straight line that is parallel to the $\lambda$-axis. 
In the region above the $d_1=d_2$ surface, where $d_1$ is less than $d_2$ and remains constant, the QD of $\rho_{\rm BD}^{\rm deph}$ remains unchanged. Conversely, in the region below the $d_1=d_2$ surface, where $d_2$ is smaller than $d_1$ and decays as $\lambda$ approaches 0, the QD of $\rho_{\rm BD}^{\rm deph}$ gradually decreases. 
Therefore, the QD dynamics can exhibit a sudden transition whenever its trajectory intersects the $d_1=d_2$ surface.

\begin{figure}
\centering
\includegraphics[width=1\columnwidth]{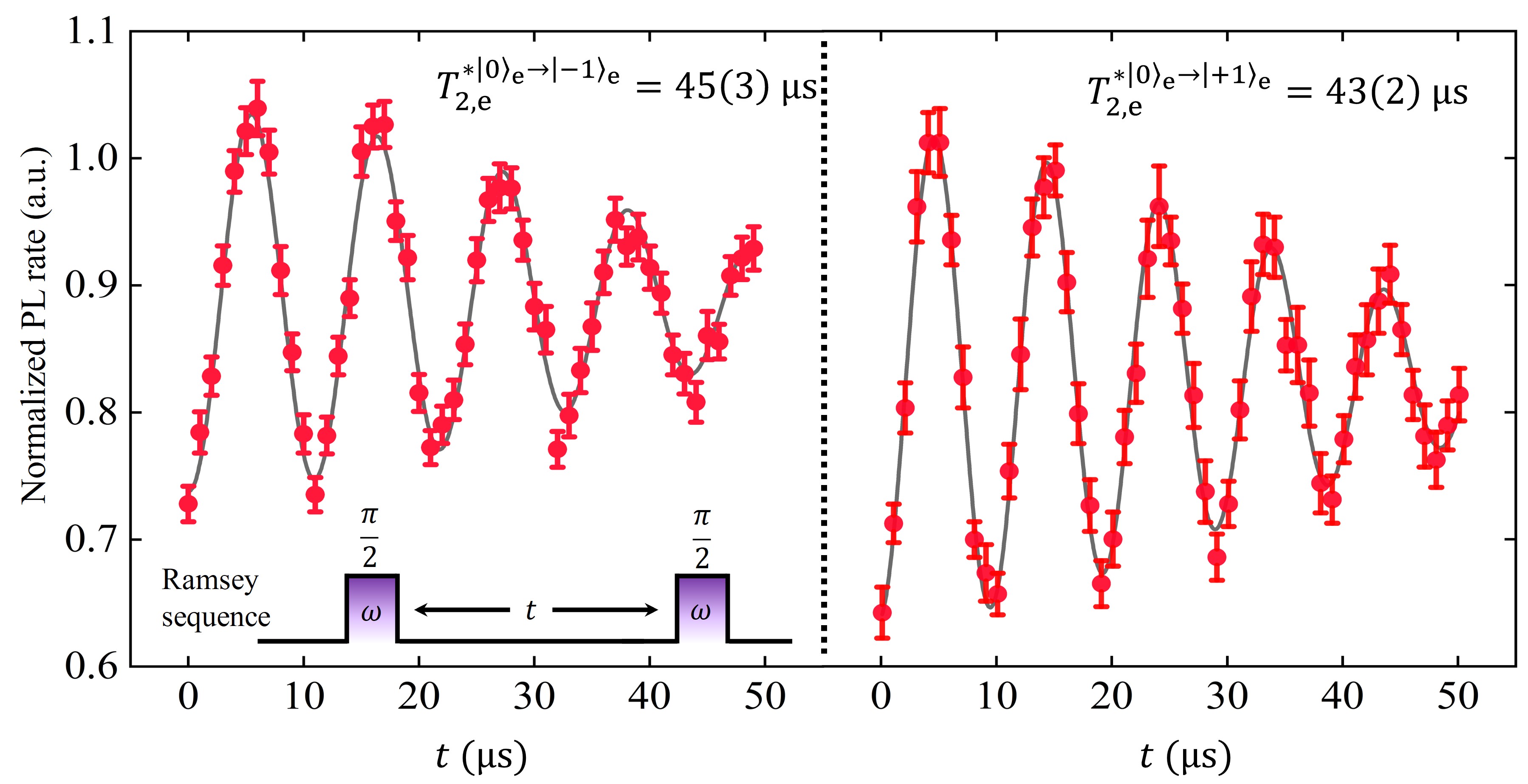}
\caption{\textbf{ Dephasing times of the electron spin qutrit.} The dephasing time for the transition $|0\rangle_{\rm e}\to |-1\rangle_{\rm e}$ ($|0\rangle_{\rm e}\to |+1\rangle_{\rm e}$) within the subspace of the nuclear spin state $|+1\rangle_
{\rm n}$ was measured via the Ramsey sequence. 
Corresponding $\omega=\omega_{|0\rangle_{\rm e}\to|- 1\rangle_{\rm e}}+\delta$ ( $\omega=\omega_{|0\rangle_{\rm e}\to|+ 1\rangle_{\rm e}}+\delta$), respectively, where $\delta=10$ kHz represents the detuning of the MW pulse. 
The Rabi frequency of these MW pulses was calibrated to 0.2 MHz.}
\label{electronSpin}
\end{figure}

\section{Preparation and Measurement of the  Bell-Diagonal State}

\subsection{Experiment setup}

The experimental setup consists of three parts: the microwave system, the optical system and the sample.

The microwave system generates and transmits microwave (MW) and radio-frequency (RF) pulses to manipulate the NV center.
The MW pulses used in our experiment were generated by an arbitrary wave generator (Keysight M8190A), amplified by an amplifier (Mini-Circuits, ZVE-3W-183+) before being fed into a diplexer (Marki DPX0R5+DPX-0508).
The RF pulses were generated by another port of the arbitrary wave generator, amplified by an amplifier (Mini Circuit LZY-22+), and then fed into the diplexer.
Finally, the MW and RF pulses combined by the diplexer were fed into a home-designed coplanar waveguide to manipulate the evolution of the NV center.

The optical system contains two components, the optic pumping component and the fluorescence collection component.
The pumping component generates 532-nm laser pulses to initialize and readout the spin state of the NV center. 
A polarizing beam splitter (PBS121) selects the S-polarized beam of the continuous 532-nm laser (generated by MSL-III-532-150mW).
The selected beam went through an acousto-optic modulator (ISOMET, AOMO 3200-121) twice to obtain 532-nm laser pulses and to decrease the laser leakage.
A quarter-wave plate (WPQ05ME-532) together with a mirror was utilized to change the polarization and transmission direction of the laser pulses so they can pass through the acousto-optic modulator twice.
Afterward, the laser pulses were coupled into an optical fiber via a reflective collimator (F810FC-543) after the expansion of a beam expander (GBE05-A).
The 532-nm laser pulses were reflected by a dichroic mirror and a mirror.
Finally, the laser pulses were focused to the sample by an oil object (Olympus, PLAPON 60*O, NA 1.45).
The fluorescence emitted by the sample went through the same oil object and was collected by an avalanche photodiode (Perkin Elmer, SPCM-AQRH-14). 
The photon counting was processed by a counter card.

\begin{figure}
\centering
\includegraphics[width=1\columnwidth]{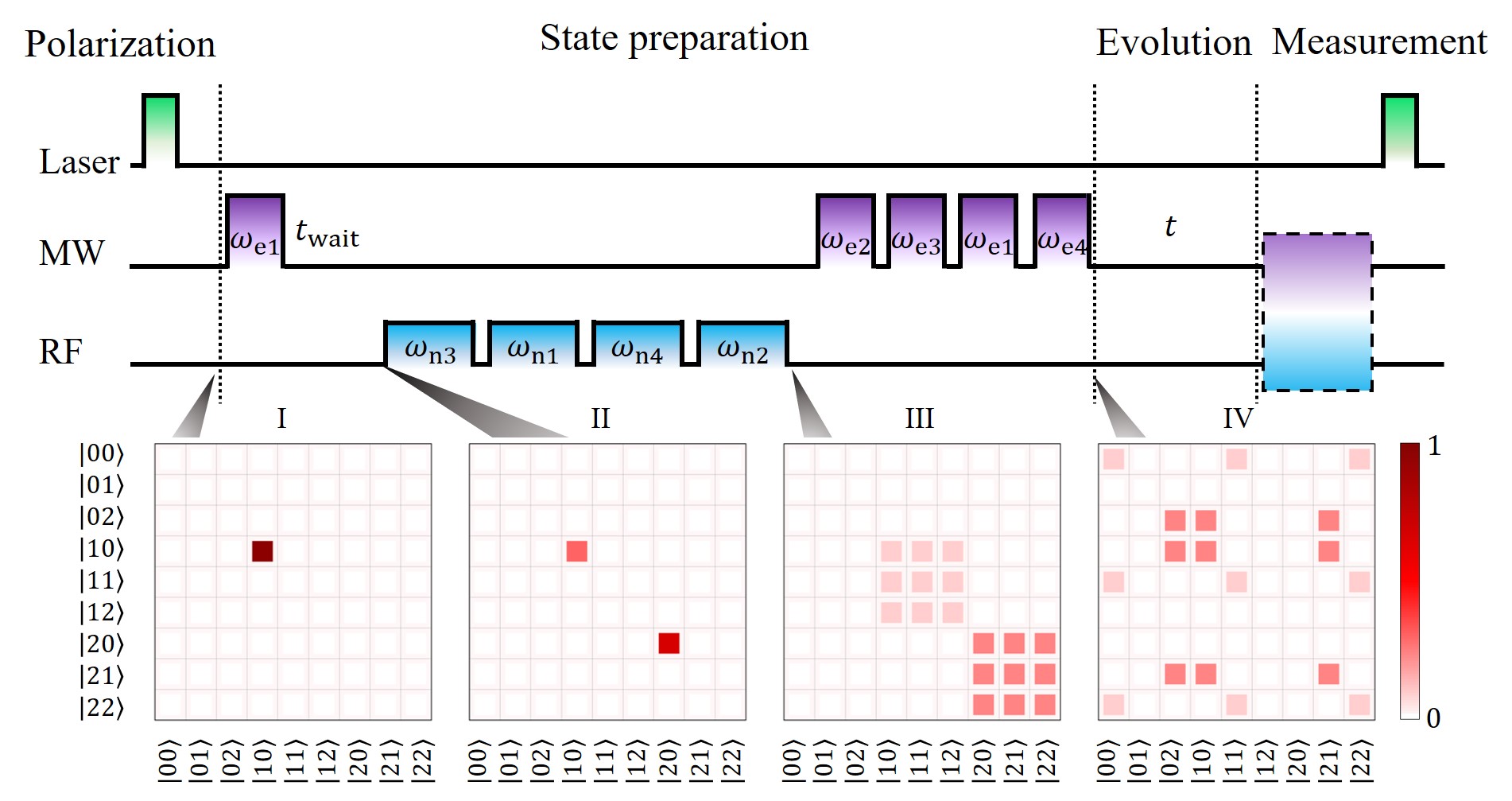}
\caption{\textbf{ Diagram of the pulse sequence.} Four parts, polarization, state preparation, evolution under the dephasing noise, and measurement are included. Information of the pulses is concluded in table \ref{state praparation parameters}. The matrices below show the theoretical evolution of the two-qutrit system. }
\label{Pulse_Sequence}
\end{figure}

The diamond was placed on a homemade coplanar waveguide in the confocal setup. 
The static magnetic field was provided by a cylindrical permanent magnet ($H=D=3$ cm).
A three-axis stage was utilized to adjust the position of the magnet, such that the direction of the magnetic field is along the symmetry axis of the NV center.
In our experiment, the static magnetic field was $B\approx 500\ {\rm G}$ to effectively polarize the electron and nuclear spin with 0.92(1) and 0.98(1) polarization rates, respectively.

The diamond sample utilized in this work was isotopically purified
with the concentration of the $^{12}{\rm C}$ atom exceeding 99.9\%. 
The dephasing times of the electron spin qutrit were measured and the results are presented in Fig.\ref{electronSpin}.
We fitted the data to the function $A~{\rm exp}\left[-(t/T_{2,\rm e}^*)^n\right]\sin[\omega' (t-t_0)] +y_0$.
The results is $T_2^*=45(3)~\mu s$ for transition $|0\rangle_{\rm e}\to|-1\rangle_{\rm e}$ and $T_2^*=43(2)~\mu s$ for transition $|0\rangle_{\rm e}\to|+1\rangle_{\rm e}$.

\begin{table*}[!htbp]
\caption{\textbf{ Detailed information about the parameters of the MW and RF pulses used in the process of the Bell-diagonal state preparation.}}
\begin{tabular}{p{1.5cm}<{\centering}|c|p{4cm}<{\centering}|p{5cm}<{\centering}}
\toprule
step & pulse frequency & transition states & pulse phase and rotation angle \\
\midrule
\multirow{1}{* }{$\rm I$} &  $ \omega_{\rm e1}$ & $|10\rangle\leftrightarrow |20\rangle$      & Y ,  $\alpha=2\arccos \sqrt{c_0}$ \\
\midrule
\multirow{4}{*}{$\rm II$} & $ \omega_{\rm n3}$ & $|10\rangle\leftrightarrow |11\rangle$    & Y ,  $\theta_1=2\arccos \sqrt{\frac{1}{3}}$ \\
 & $  \omega_{\rm n1}$ & $|20\rangle\leftrightarrow |21\rangle$    & Y ,  $\theta_2=2\arccos \sqrt{\frac{1}{3}}$ \\
&$  \omega_{\rm n4}$ & $|11\rangle\leftrightarrow |12\rangle$      & Y ,  $\theta_3=2\arccos \sqrt{\frac{1}{2}}$ \\
&$  \omega_{\rm n2}$ & $|21\rangle\leftrightarrow |22\rangle$      & Y ,  $\theta_4=2\arccos \sqrt{\frac{1}{2}}$ \\
\midrule
\multirow{4}{*}{$\rm III$}&$\omega_{\rm e2}$ & $|00\rangle\leftrightarrow |10\rangle$      & -Y ,  $\pi$ \\
&$ \omega_{\rm e3}$ & $|12\rangle\leftrightarrow |22\rangle$      & Y ,  $\pi$ \\
&$ \omega_{\rm e1}$ & $|10\rangle\leftrightarrow |20\rangle$    & -Y ,  $\pi$ \\
&$\omega_{\rm e4}$ & $|02\rangle\leftrightarrow |12\rangle$    & Y ,  $\pi$ \\
\bottomrule
\end{tabular}
\label{state praparation parameters}
\end{table*}

\subsection{Preparation of the Bell-Diagonal State}

The two cases of the Bell-diagonal states investigated in the experiment have the form $\rho_{\rm BD} = c_0|\Psi_{00}\rangle \langle\Psi_{00}| + c_1|\Psi_{01}\rangle\langle\Psi_{01}| + c_2|\Psi_{02}\rangle\langle \Psi_{02}|$ with restrictions $c_0+c_1+c_2=1$ and $c_1=0$. The corresponding density matrix can be written as
\begin{equation}
\rho_{\rm BD}^{\rm exp} =\frac{1}{3}
\begin{pmatrix}
c_0& 0 & 0 & 0 &c_0& 0 & 0 & 0 & c_0\\
 0 & 0 & 0 & 0 & 0 & 0 & 0 & 0 & 0\\
 0 & 0 &c_2&c_2& 0 & 0 & 0 &c_2& 0\\
 0 & 0 &c_2&c_2& 0 & 0 & 0 &c_2& 0\\
c_0& 0 & 0 & 0 &c_0& 0 & 0 & 0 & c_0\\
 0 & 0 & 0 & 0 & 0 & 0 & 0 & 0 & 0\\
 0 & 0 & 0 & 0 & 0 & 0 & 0 & 0 & 0\\
 0 & 0 &c_2&c_2& 0 & 0 & 0 & c_2 & 0\\
c_0& 0 & 0 & 0 &c_0& 0 & 0 & 0 & c_0
\end{pmatrix} .
\end{equation}
The electron (nuclear) spin corresponds to the first (second) qutrit and the states $\rm |-1\rangle_e,|0\rangle_e,|+1\rangle_e$ ($\rm |-1\rangle_n,|0\rangle_n,|+1\rangle_n$) correspond to the $|0\rangle,|1\rangle,|2\rangle$ of the first (second) qutrit, respectively. MW and radio frequency (RF) pulses were utilized to control the state of the electron spin and the nuclear spin, respectively. 

\begin{figure}
\centering
\includegraphics[width=0.7\columnwidth]{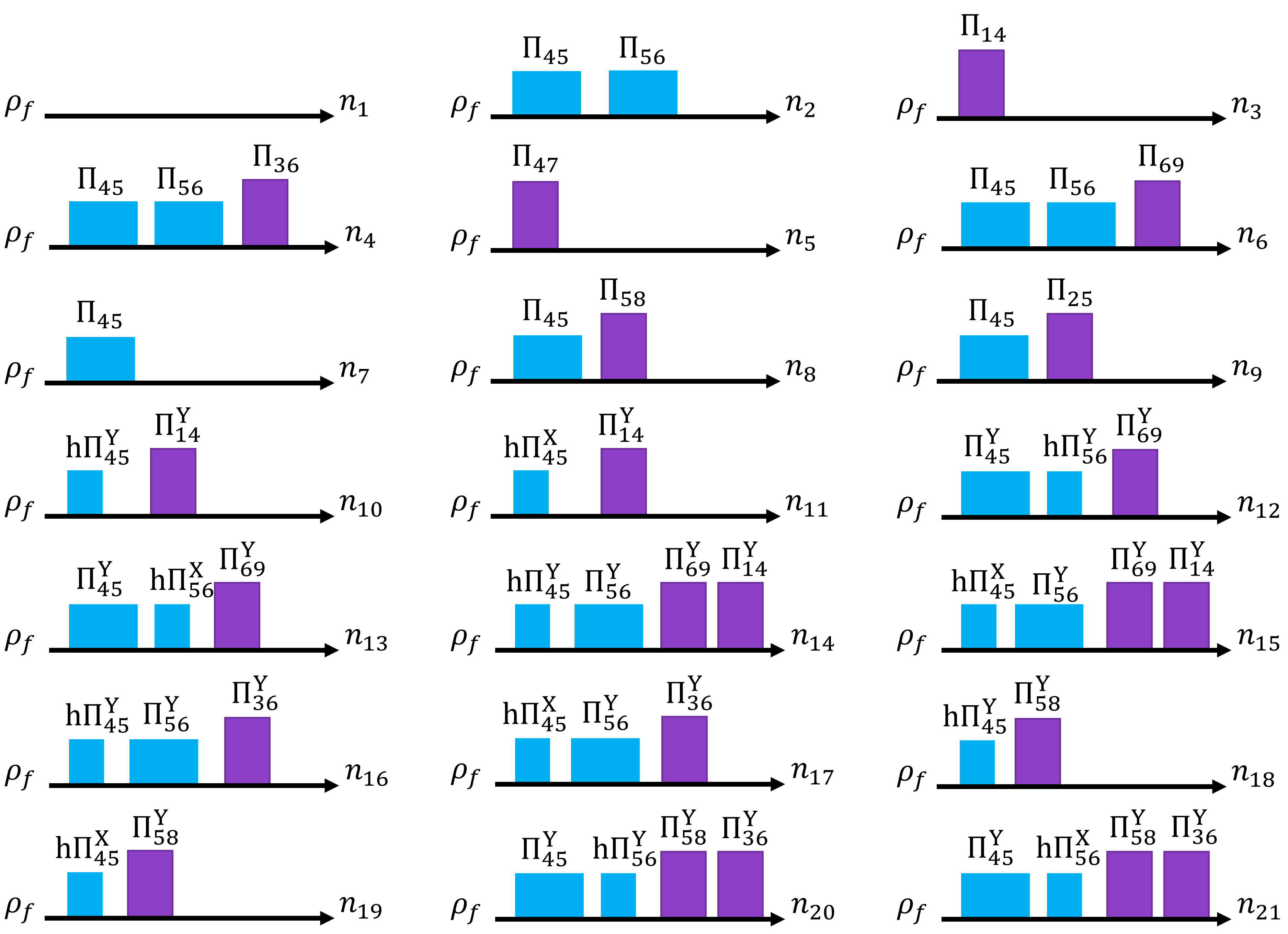}
\caption{\textbf{Measurement pulse sequences of the Bell-diagonal state.}
$\rho_f$ denotes the density matrix of the prepared Bell-diagonal state.
${\rm \Pi}_{m,n}^{\rm X(Y)}$ (${\rm h}{\rm \Pi}_{m,n}^{\rm X(Y)}$) denotes the selective $\pi$ ($\pi/2$) pulse applied between state $|m\rangle$ and state $|n\rangle$ along X (Y) axis.
$n_j$ ($j\in[1,21]$) indicates the detected photoluminescence rate after applying these pulse sequences.
}
\label{tomography sequences}
\end{figure}

As shown in Fig. \ref{Pulse_Sequence}, these states can be prepared following the procedures below.
\begin{itemize}
\item[(I)]
Apply a 532-nm laser pulse to polarize the two-qutrit system to state $|10\rangle$, as shown by the first density matrix in Fig. \ref{Pulse_Sequence}.
\item[(II)]
Apply a selective MW pulse with frequency $\omega_{\rm e1}$ to transfer the polarized state to state $\sqrt{c_0}|{10}\rangle +\sqrt{c_2}|20\rangle$. Then, waiting a free evolution time of 200 $\mu$s, the coherence of the electron spin will be dissipated. The state will be left with the form $\rho_{\rm II}=c_0|10\rangle\langle10| + c_2|20\rangle\langle20|$ as shown by the second-density matrix in Fig. \ref{Pulse_Sequence}.
\item[(III)]
Apply four selective RF pulses of frequency $\omega_{\rm n3}$, $\omega_{\rm n1}$, $\omega_{\rm n4}$ and $\omega_{\rm n2}$ with appropriate time-duration in sequence.
The system will be prepared to state $\rho_{\rm II}$ as shown by the third-density matrix in Fig. \ref{Pulse_Sequence},
\begin{equation}
\rho_{\rm II} =\frac{1}{3}
\begin{pmatrix}
0 & 0 & 0 & 0 & 0 & 0 & 0 & 0 & 0\\
0 & 0 & 0 & 0 & 0 & 0 & 0 & 0 & 0\\
0 & 0 & 0 & 0 & 0 & 0 & 0 & 0 & 0\\
0 & 0 & 0 & c_0 & c_0 & c_0 & 0 & 0 & 0\\
0 & 0 & 0 & c_0 & c_0 & c_0 & 0 & 0 & 0\\
0 & 0 & 0 & c_0 & c_0 & c_0 & 0 & 0 & 0\\
0 & 0 & 0 & 0 & 0 & 0 & c_2 & c_2 & c_2\\
0 & 0 & 0 & 0 & 0 & 0 & c_2 & c_2 & c_2\\
0 & 0 & 0 & 0 & 0 & 0 & c_2 & c_2 & c_2
\end{pmatrix} \ .
\end{equation}
\item[(IV)]
Apply four selective MW $\pi$ pulses with frequency $\omega_{\rm e2}$, $\omega_{\rm e3}$, $\omega_{\rm e1}$ and $\omega_{\rm e4}$ in sequence. The system will finally be prepared to $\rho_{\rm BD}^{\rm exp}$ as shown by the last density matrix in Fig. \ref{Pulse_Sequence}.
\end{itemize}

The Rabi frequency of the MW (RF) pulses was all set to 0.2 MHz (25 kHz). Details about which two energy levels these pulses worked on, the phase, and the time duration (denoted by the rotation angle) of these pulses are given in the table \ref{state praparation parameters} below. Utilizing this method, the initial Bell-diagonal states in the blue and red trajectory of Figure 1 in the main text were prepared with fidelity 0.96(0.01) and 0.95(0.01), respectively.

\subsection{Measurement of the Dephased Bell-Diagonal States}

The final state of the two-qutrit system after an evolution of time duration $\tau$ was measured via a set of pulse sequences. 
It is noticed that we did not measure all the elements in the density matrix to reduce the experimental complexity. 
Specifically, elements that are non-zero theoretically were all measured. 
Whereas, for elements that are zero theoretically, we only measured a part of them. Experimental results of these elements were very close to zero, showing this compromise is acceptable. The pulse sequences to measure the non-zero elements of the density matrix were given in Fig. \ref{tomography sequences}, where we relabeled $|00\rangle$, $|01\rangle$, $|02\rangle$, $|10\rangle$, $|11\rangle$, $|12\rangle$, $|20\rangle$, $|21\rangle$, and $|22\rangle$ as $|1\rangle$, $|2\rangle$, $|3\rangle$, $|4\rangle$, $|5\rangle$, $|6\rangle$, $|7\rangle$, $|8\rangle$, and $|9\rangle$ for simplicity. With a normalized photoluminescence of different energy levels, all these elements can be calculated.

 We employed a maximum likelihood estimation (MLE) method to acquire the most possible physical state $\rho_{\rm MLE}$ \cite{PRA_2001_James}.
After that, the QD of the $\rho_{\rm MLE}$ was obtained. Details about the maximum likelihood estimation and the calculation of the QD were introduced in Ref. \onlinecite{PRL_2022_Fu}.


\begin{thebibliography}{99}
	
	
	\bibitem{review_2009} R. Horodecki, P. Horodecki, M. Horodecki, and K. Horodecki, Quantum Entanglement, \emph{Rev. Mod. Phys.} \textbf{81}, 865 (2009).	
	\bibitem{review_2020_Uola} R. Uola, A. C. S. Costa, H. C. Nguyen, and O. Gühne, Quantum Steering, Rev. Mod. Phys. 92, 015001 (2020).	
	\bibitem{book_2015_streltsov} A. Streltsov, Quantum Correlations Beyond Entanglement and Their Role in Quantum Information Theory (Springer, New York, 2015).
	\bibitem{JPhys_2016_Adesso} G. Adesso, T. R. Bromley, and M. Cianciaruso, Measures and Applications of Quantum Correlations, \emph{ J. Phys. A: Math. Theor.} \textbf{49}, 473001 (2016).
	
	\bibitem{review_2018_Braun} D. Braun, G. Adesso, F. Benatti, R. Floreanini, U. Marzolino, M. W. Mitchell, and S. Pirandola, Quantum-Enhanced Measurements without Entanglement, \emph{Rev. Mod. Phys.} \textbf{90}, 035006 (2018).	
	
	
	\bibitem{review_2016_Suter} D. Suter and G. A. Álvarez, Colloquium: Protecting Quantum Information against Environmental Noise, \emph{Rev. Mod. Phys.} \textbf{88}, 041001 (2016).
	
	\bibitem{review_2019_Lewis-Swan} R. J. Lewis-Swan, A. Safavi-Naini, A. M. Kaufman, and A. M. Rey, Dynamics of Quantum Information, \emph{Nat. Rev. Phys.} \textbf{1}, 627 (2019).
	
	\bibitem{NC_2021_wang}P. Wang, C.-Y. Luan, M. Qiao, M. Um, J. Zhang, Y. Wang, X. Yuan, M. Gu, J. Zhang, and K. Kim, Single ion qubit with estimated coherence time exceeding one hour, \emph{Nat. Commun.} \textbf{12}, 233 (2021).
	
	\bibitem{PRL_2023_zhang}C. Zhang, P. Yu, A. Jadbabaie, and N. R. Hutzler, Quantum-enhanced metrology for molecular symmetry violation using decoherence-free subspaces, \emph{Phys. Rev. Lett.} \textbf{131}, 193602 (2023).
	
	\bibitem{PRL_2024_wang}H.-R. Wang, D. Yuan, S.-Y. Zhang, Z. Wang, D.-L. Deng, and L. M. Duan, Embedding quantum many-body scars into decoherence-free subspaces, \emph{Phys. Rev. Lett.} \textbf{132}, 150401 (2024).
	
	\bibitem{NC_2023_Sundaresan} N. Sundaresan, T. J. Yoder, Y. Kim, M. Li, E. H. Chen, G. Harper, T. Thorbeck, A. W. Cross, A. D. Corcoles, and M. Takita, A. D. C$\acute{o}$rcoles, and M. Takita, Demonstrating multi-round subsystem quantum error correction using matching and maximum likelihood decoders. \emph{Nat. Commun.} \textbf{14}, 2852 (2023).
	\bibitem{PRL_2010_Mazzola} L. Mazzola, J. Piilo, and S. Maniscalco, Sudden Transition between Classical and Quantum Decoherence, \emph{Phys. Rev. Lett.} \textbf{104}, 200401 (2010).
	\bibitem{PRA_2011_Karpat}G. Karpat, Z. Gedik, Correlation dynamics of qubit-qutrit systems in a classical dephasing environment. Phys. Lett. A 375, 4166 (2011)
	
	\bibitem{PRL_2015_Carnio}E.G. Carnio, A. Buchleitner, M. Gessner, Robust asymptotic entanglement under multipartite collective dephasing. Phys. Rev. Lett. 115, 010404 (2015)
	
	\bibitem{PRL_2001_Olivier} H. Ollivier and W. H. Zurek, Quantum Discord: A Measure of the Quantumness of Correlations, \emph{Phys. Rev. Lett.} \textbf{88}, 017901 (2001).
	
	\bibitem{JPA_2001_Henderson} L. Henderson and V. Vedral, Quantum and Total Correlations, \emph{J. Phys. A: Math. Gen.} \textbf{34}, 6899 (2001).
	
	\bibitem{review_2012_Modi} K. Modi, A. Brodutch, H. Cable, T. Paterek, and V. Vedral, The Classical-Quantum Boundary for Correlations: Discord and Related Measures, \emph{Rev. Mod. Phys.} \textbf{84}, 1655 (2012).
	
	\bibitem{review_2018_Bera}A. Bera, T. Das, D. Sadhukhan, S. Singha Roy, A. Sen(De), and U. Sen, Quantum discord and its allies: a review of recent progress. \emph{Rep. Prog. Phys.} \textbf{81}, 024001 (2018).
	
	
	
	\bibitem{review_2018_De} G. De Chiara and A. Sanpera, Genuine Quantum Correlations in Quantum Many-Body Systems: A Review of Recent Progress, \emph{Rep. Prog. Phys.} \textbf{81}, 074002 (2018).
	
	
	\bibitem{NP_2012_Dakić}B. Daki$\acute{c}$, Y. O. Lipp, X. Ma, M. Ringbauer, S. Kropatschek, S. Barz, T. Paterek, V. Vedral, A. Zeilinger, Č. Brukner et al., Quantum Discord as Resource for Remote State Preparation, \emph{Nat. Phys.} \textbf{8}, 666 (2012).	
	
	
	
	\bibitem{NC_2013_Xu}J.-S. Xu, K. Sun, C.-F. Li, X.-Y. Xu, G.-C. Guo, E. Andersson, R. Lo Franco, and G. Compagno, Experimental Recovery of Quantum Correlations in Absence of System-Environment Back-Action, \emph{Nat. Commun.} \textbf{4}, 2851 (2013).
	
	
	\bibitem{PRL_2011_Auccaise} R. Auccaise, L. C. C$\rm \acute{e}$leri, D. O. Soares-Pinto, E. R. deAzevedo, J. Maziero, A. M. Souza, T. J. Bonagamba, R. S. Sarthour, I. S. Oliveira, and R. M. Serra, Environment-Induced Sudden Transition in Quantum Discord Dynamics, \emph{Phys. Rev. Lett.} \textbf{107}, 140403 (2011).
	\bibitem{NC_2010_Xu} J.-S. Xu, X.-Y. Xu, C.-F. Li, C.-J. Zhang, X.-B. Zou, and G.-C. Guo, Experimental Investigation of Classical and Quantum Correlations under Decoherence, \emph{Nat. Commun.} \textbf{1}, 7 (2010).
	\bibitem{EPL_2017_Singh} H. Singh, Arvind, and K. Dorai, Experimentally Freezing Quantum Discord in a Dephasing Environment Using Dynamical Decoupling, \emph{EPL} \textbf{118}, 50001 (2017).
	
	
	\bibitem{AQT_2019_Cozzolino} D. Cozzolino, B. Da Lio, D. Bacco, and L. K. Oxenlwe, High‐Dimensional Quantum Communication: Benefits, Progress, and Future Challenges, \emph{Adv. Quantum Technol.} \textbf{2}, 1900038 (2019).
	
	\bibitem{FP_2020_Wang} Y. Wang, Z. Hu, B. C. Sanders and S. Kais, Qudits and High-Dimensional Quantum Computing, \emph{Front. Phys.} \textbf{8}, 479 (2020).
	\bibitem{review_2020_Erhard}M. Erhard, M. Krenn, and A. Zeilinger, Advances in High-Dimensional Quantum Entanglement, \emph{Nat. Rev. Phys.}\textbf{2}, 365 (2020).
	
	
	\bibitem{review_2023_Guo} X.-M. Hu, Y. Guo, B.-H. Liu, C.-F. Li, and G.-C. Guo, Progress in Quantum Teleportation, \emph{Nat. Rev. Phys.} \textbf{5}, 339 (2023).
	
	
	\bibitem{NC_2023_Hrmo}P. Hrmo, B. Wilhelm, L. Gerster, M. W. Van Mourik, M. Huber, R. Blatt, P. Schindler, T. Monz, and M. Ringbauer, Native Qudit Entanglement in a Trapped Ion Quantum Processor, \emph{Nat. Commun.} \textbf{14}, 2242 (2023).
	\bibitem{Science_2018_Wang}  J. Wang, S. Paesani, Y. Ding, R. Santagati, P. Skrzypczyk, A. Salavrakos, J. Tura, R. Augusiak, L. Mančinska, D. Bacco et al., Multidimensional Quantum Entanglement with Large-Scale Integrated Optics, \emph{Science} \textbf{360}, 285 (2018).
	
	
	
	\bibitem{PRX_2023_Liu} Pei Liu, Ruixia Wang, Jing-Ning Zhang, Yingshan Zhang,
	Xiaoxia Cai, Huikai Xu, Zhiyuan Li, Jiaxiu Han, Xuegang Li, Guangming Xue et al., Performing SU($d$) Operations and Rudimentary Algorithms in a Superconducting Transmon Qudit for $d = 3$ and $d = 4$, \emph{Phys. Rev. X} \textbf{13}, 021028 (2023).
	\bibitem{NC_2024_Fernández} I. F. de Fuentes, T. Botzem, M. A. I. Johnson, A. Vaartjes, S. Asaad, V. Mourik, F. E. Hudson, K. M. Itoh, B. C. Johnson,
	A. M. Jakob et al., Navigating the 16-Dimensional Hilbert Space of a High-Spin Donor Qudit with Electric and Magnetic Fields, \emph{Nat. Commun.} \textbf{15}, 1380 (2024).
	
	\bibitem{NC_2022_Chi} Y. Chi, J. Huang, Z. Zhang, J. Mao, Z. Zhou, X. Chen, C. Zhai, J. Bao, T. Dai, H. Yuan et al., A Programmable Qudit-Based Quantum Processor, \emph{Nat. Commun.} \textbf{13}, 1166 (2022).
	\bibitem{NP_2022_Ringbauer}M. Ringbauer, M. Meth, L. Postler, R. Stricker, R. Blatt, P. Schindler, and T. Monz, A Universal Qudit Quantum Processor with Trapped Ions, \emph{Nat. Phys.} \textbf{18}, 1053 (2022).
	\bibitem{NP_2019_Reimer} C. Reimer, S. Sciara, P. Roztocki, M. R. Islam, L. Cortés, Y.
	Zhang, B. Fischer, S. Loranger, R. Kashyap, A. Cino et al., High-Dimensional One-Way Quantum Processing Implemented on $d$-Level Cluster States \emph{Nat. Phys.} \textbf{15}, 148-153 (2019).
	
	\bibitem{PRA_2015_Vitanov}N. V. Vitanov, Dynamical rephasing of ensembles of qudits. \emph{Phys. Rev. A} \textbf{92}, 022314 (2015).
	
	
	\bibitem{NC_2020_Coladangelo} A. Coladangelo and J. Stark, An Inherently Infinite-Dimensional Quantum Correlation, \emph{Nat. Commun.} \textbf{11}, 3335 (2020).
	
	
	\bibitem{PRR_2021_Napolitano} R. de J. Napolitano, F. F. Fanchini, A. H. da Silva, and B. Bellomo, Protecting operations on qudits from noise by continuous dynamical decoupling. \emph{Phys. Rev. Research} \textbf{3}, 013235 (2021).
	
	
	
	
	\bibitem{PS_2022_Singh}A. Singh and U. Sinha, Entanglement protection in higher-dimensional systems. \emph{Phys. Scr.} \textbf{97}, 085104 (2022).
	
	
	
	
	\bibitem{PRL_2018_Kraft}T. Kraft, C. Ritz, N. Brunner, M. Huber, and O. Gühne, Characterizing Genuine Multilevel Entanglement, \emph{Phys. Rev. Lett.} \textbf{120}, 060502 (2018).
	
	
	\bibitem{PRA_2022_yuan}Xinxing Yuan, Yue Li, Mengxiang Zhang, Chang Liu,
	Mingdong Zhu, Xi Qin, Nikolay V. Vitanov, Yiheng Lin, and Jiangfeng Du, Preserving multilevel quantum coherence by dynamical decoupling. \emph{Phys. Rev. A} \textbf{106}, 022412 (2022).
	
	
	
	
	\bibitem{PRL_2022_Fu}Y. Fu, W. Liu, X. Ye, Y. Wang, C. Zhang, C.-K. Duan, X. Rong, and J. Du, Experimental Investigation of Quantum Correlations in a Two-Qutrit Spin System, \emph{Phys. Rev. Lett.} \textbf{129}, 100501 (2022).
	
	
	
	
	\bibitem{PRA_2010_Ali} M. Ali, Distillability Sudden Death in Qutrit-Qutrit Systems under Global and Multilocal Dephasing, \emph{Phys. Rev. A} \textbf{81}, 042303 (2010).
	
	\bibitem{IJQI_2022_xiao} W. Xiao, M.-Y. Zhen, and X.-W. Hou, Freezing of Geometric Discords for Two Qutrits in Environments, \emph{Int. J. Quantum Inform.} \textbf{20}, 2250019 (2022).
	\bibitem{SR_2017_Cárdenas} F. A. Cárdenas-López, S. Allende, and J. C. Retamal, Sudden Transition between Classical to Quantum Decoherence in Bipartite Correlated Qutrit Systems, \emph{Sci. Rep.} \textbf{7}, 44654 (2017).
	
	
	
	
	
	\bibitem{PRA_2009_Maziero} J. Maziero, L. C. Céleri, R. M. Serra, and V. Vedral, Classical and Quantum Correlations under Decoherence, \emph{Phys. Rev. A} \textbf{80}, 044102 (2009).
	
	\bibitem{PRA_2009_Werlang}T. Werlang, S. Souza, F. F. Fanchini, and C. J. Villas Boas, Robustness of Quantum Discord to Sudden Death, \emph{Phys. Rev. A} \textbf{80}, 024103 (2009).
	%
	\bibitem{PRL_2010_Lang} M. D. Lang and C. M. Caves, Quantum Discord and the Geometry of Bell-Diagonal States, \emph{Phys. Rev. Lett.} \textbf{105}, 150501 (2010).
	{\bibitem{PRA_2015_Chanda} T. Chanda, A. K. Pal, A. Biswas, A. Sen(De), and U. Sen, Freezing of Quantum Correlations under Local Decoherence, \emph{Phys. Rev. A} \textbf{91}, 062119 (2015).}
	
	
	
	
	
	
	
	
	
	
	
	
	
	
	
	\bibitem{PhysRep_2013_Doherty} M. W. Doherty, N. B. Manson, P. Delaney, F. Jelezko, J. Wrachtrup, and L. C. L. Hollenberg, The Nitrogen-Vacancy Colour Centre in Diamond, \emph{Phys. Rep.} \textbf{528}, 1 (2013).
	
	\bibitem{Review_2017_Cham}Lectures on General Quantum Correlations and Their Applications. (Springer International Publishing, Cham, 2017). 
	
	\bibitem{PRL_2010_Dakić} B. Dakić, V. Vedral, and Č. Brukner, Necessary and Sufficient Condition for Nonzero Quantum Discord, \emph{Phys. Rev. Lett.} \textbf{105}, 190502 (2010).
	\bibitem{EPL_2013_Paula} F. M. Paula, J. D. Montealegre, A. Saguia, T. R. De Oliveira, and M. S. Sarandy, Geometric Classical and Total Correlations via Trace Distance, \emph{Europhys. Lett.} \textbf{103}, 50008 (2013).
	\bibitem{PLA_2016_Jakóbczyk} L. Jakobczyk, A. Frydryszak, and P. Lugiewicz, \emph{Phys. Lett. A} \textbf{380}, 1535 (2016).
	
	
	
	
	\bibitem{review_2021_Wolfowicz}G. Wolfowicz, F. J. Heremans, C. P. Anderson, S. Kanai, H. Seo, A. Gali, G. Galli, and D. D. Awschalom, Quantum Guidelines for Solid-State Spin Defects, \emph{Nat. Rev. Mater.} \textbf{6}, 906 (2021).
	
	
	
	
	
	
	\bibitem{PRL_2009_Jacques} V. Jacques, P. Neumann, J. Beck, M. Markham, D. Twitchen, J. Meijer, F. Kaiser, G. Balasubramanian, F. Jelezko, and J. Wrachtrup, Dynamic Polarization of Single Nuclear Spins by Optical Pumping of Nitrogen-Vacancy Color Centers in Diamond at Room Temperature, \emph{Phys. Rev. Lett.} \textbf{102}, 057403 (2009).
	
	\bibitem{PRA_2001_James}D. F. V. James, P. G. Kwiat, W. J. Munro, and A. G. White, Measurement of Qubits, \emph{Phys. Rev. A} \textbf{64}, 052312 (2001).
	
	
	
	\bibitem{PRA_2012_Rossignoli} R. Rossignoli, J. M. Matera, and N. Canosa, Measurements, Quantum Discord, and Parity in Spin-1 Systems, \emph{Phys. Rev. A 86} \textbf{473}, 022104 (2012). 
	
	\bibitem{PRA_2007_Derkacz}L. Derkacz and L. Jakobczyk, Entanglement versus Entropy for a Class of Mixed Two-Qutrit States, \emph{Phys. Rev. A} \textbf{76}, 042304 (2007).
	
	
	
	
	
	{\bibitem{PRB_2013_Fischer} R. Fischer, A. Jarmola, P. Kehayias, and D. Budker, Optical Polarization of Nuclear Ensembles in Diamond, \emph{Phys. Rev. B} \textbf{87}, 125207 (2013).}
	
	
	
	
	
	
	
	
	%
	%
	
	
	
	
	
	
	
	
\end{thebibliography}
\end{document}